\documentclass[aip,amsmath,amssymb,graphicx,reprint]{revtex4-2}

\usepackage{graphicx}
\usepackage{cancel}
\usepackage{dcolumn}
\usepackage{xcolor}

\draft 

\begin{document}

\title{Impact of the ligand deformation on the $\mathcal{P}$,$\mathcal{T}$-violation effects in the YbOH molecule}

\author{Anna Zakharova} \email{zakharova.annet@gmail.com}
\affiliation{St. Petersburg State University, St. Petersburg, 7/9 Universitetskaya nab., 199034, Russia} 
\affiliation{Petersburg Nuclear Physics Institute named by B.P. Konstantinov of National Research Centre
"Kurchatov Institute", Gatchina, 1, mkr. Orlova roshcha, 188300, Russia}

\author {Alexander Petrov}\email{petrov\_an@pnpi.nrcki.ru}

\affiliation{St. Petersburg State University, St. Petersburg, 7/9 Universitetskaya nab., 199034, Russia} 
\affiliation{Petersburg Nuclear Physics Institute named by B.P. Konstantinov of National Research Centre
"Kurchatov Institute", Gatchina, 1, mkr. Orlova roshcha, 188300, Russia}

\date{\today}

\begin{abstract}
The ytterbium monohydroxide is a promising molecule for a new physics searches. It is well known that levels of the opposite parity,  separated by the energy split, so-called $l$-doublets, define the experimental electric field strength required for the molecule polarization. In addition, in our previous paper \cite{petrov2022sensitivity} we have shown that the value of $l$-doubling directly influences the sensitivity of linear triatomic molecules to the $\mathcal{P}$,$\mathcal{T}$-odd effects.
In our work \cite{zakharova2021rovibrational} we have calculated the value of $l$-doubling for the YbOH molecule with approximation of fixed O-H bond length.
Accounting the importance of this property, in the present study,
we consider the additional degree of freedom corresponding to the ligand (OH) deformation. 
\end{abstract}


\maketitle 
\section{Introduction}
Symmetry breaking with respect to the spatial reflection ($\mathcal{P}$), the time inversion ($\mathcal{T}$), and the charge conjugation ($\mathcal{C}$) are embedded in the Standard model (SM) \cite{schwartz2014quantum,particle2020review}. 
One of the possible sources of the strong $\mathcal{P,T}$-violation in the SM is $\theta$ term, that however is negligible \cite{Cheng1988,KimCarosi2010}.
The observed source of the $\mathcal{P}$ and $\mathcal{C}$ non-conservation comes from the weak interaction, that affects only the left components of the lepton and quark spinors. This fact means that  $\mathcal{P}$, and $\mathcal{C}$ symmetries  are broken separately \cite{khriplovich2012cp}. However, the combined $\mathcal{CP}$ parity could be conserved. But in SM there is a Cabibbo-Kobayashi-Maskawa (CKM) \cite{Cabibbo1963,KobayashiMaskawa1973} matrix, which is associated with the interaction of quarks and $W^\pm$-bosons, and Pontecorvo–Maki–Nakagawa–Sakata matrix (PMNS) \cite{Pontecorvo1957,MNS1962}, which is associated with the interaction of leptons with $W^\pm$-bosons. When these matrices have complex components, they violate $\mathcal{CP}$, which is manifested in neutral kaon and $B^0$-meson decays \cite{schwartz2014quantum}.
Because of the $\mathcal{CPT}$ theorem, $\mathcal{CP}$ combination is equivalent to time reversal $\mathcal{T}$.

The probable effect of the $\mathcal{CP}$-symmetry violation in the Standard model can be found in the electron electric dipole moment (eEDM) and scalar-pseudoscalar electron-nucleon interaction (S-PS)\cite{barr1992t}, parametrized correspondingly by $d_e$ and $k_s$ constants.
 In the SM the predicted values for these constants are very small. Nevertheless, there are many extensions of the SM that can lead to the considerable increase of the $d_e$ and $k_s$ constants \cite{Fukuyama2012,PospelovRitz2014,YamaguchiYamanaka2020,YamaguchiYamanaka2021}.

To investigate the $\mathcal{CP}$-violating physics, table-top experiments with atoms and molecules can be employed \cite{baron2014order,demille2017probing, ACME:18, cirigliano2019c, hutzler2020searches,Cornell:2017,Petrov:18, whitepaper}.
%
The current limit on the electron electric dipole moment (the ACME II experiment), $|d_e|<1.1\times 10^{-29}$ e$\cdot$cm (90\% confidence), was set by measuring the spin precession using thorium monoxide (ThO) molecules in the metastable electronic H$^3\Delta_1$ state~\cite{ACME:18}.

In its turn, cold polar molecules provide unique opportunities for further progress in search for effects of symmetry violation \cite{Isaev:16}.
The strength of polyatomic spices as probes of the parity-violating physics lies in a possibility of 
laser-cooling that does not interfere with the existence of the parity doublets \cite{Isaev_2017}.
Unlike the diatomic molecules, polyatomic ones possess the distinctive rovibrational spectrum with additional energy levels of opposite parity. The transverse vibrations of triatomic molecules result in $l$-doublets \cite{Kozyryev:17,hutzler2020polyatomic}.  For symmetric top molecules nonvibrational $K$-doublets, associated with rotation of the molecule around its axis, appear \cite{Kozyryev:17, zakharova2022rotating}.

 One dimensional laser-cooling was made for SrOH, CaO, CaOCH$_3$ \cite{kozyryev2017sisyphus,baum20201d,mitra2020direct}, and the YbOH \cite{steimle2019field,augenbraun2020laser} molecules. Recently, the magneto-optical trapping and cooling of the CaOH molecule was achieved \cite{vilas2022magneto}. The molecule we are considering, ytterbium monohydroxide, is a good candidate for the eEDM search \cite{Kozyryev:17,zakharova2021rovibrational,petrov2022sensitivity}.

For triatomic molecules, the external fields are mixing the opposite parity levels, and due to this, the molecule is polarized to a degree determined by a coefficient $P$ \cite{petrov2022sensitivity}. If the electron has eEDM and interacts with the nucleus by the S-PS, the $\mathcal{P}$, $\mathcal{T}$-nonconservation emerges in the energy difference between levels with opposite projection of the total angular momentum on the field's axis:
\begin{equation}
\Delta E_{\mathcal{P},\mathcal{T}}=P(2E_{\rm eff}  d_e + 2E_{\rm s} k_s),
\label{split}
\end{equation}
Knowing the enhancement coefficients  $E_{\rm eff}$, $E_{\rm s}$ and $P$ one may extract the value of the constants $d_e$ and $k_s$ from this energy splitting \textit{\cite{KozlovLabzowsky1995, titov2006d, Safronova2017}}. 

Recently, we have calculated the enhancement factors $E_{\rm eff}$ and $E_{\rm s}$ for the RaOH, YbOH, and the symmetric top molecule RaOCH$_3$, the rovibrational wavefunctions were obtained \cite{ourRaOH, zakharova2021rovibrational, zakharova2022rotating} on the CCSD level. Our computational approach, that does take into account the rotational and anharmonic effects, allowed us to calculate $l$--doubling for the triatomic molecules RaOH and YbOH. We have also determined the value of the polarization coefficient $P$ for the YbOH molecule \cite{petrov2022sensitivity}. 
In our work \cite{petrov2022sensitivity} we have
shown that the $l$-doubling structure is, in general, different
from $\Omega-$doubling of diatomics, and the $P$ value tends to approach 50\% value for most of the ground rotational levels of the first excited $\nu_2 = 1$
bending mode of YbOH. For some levels a maximum of the polarization
$P^{\rm max}$ value as a function of electric field with $1/2 < |P|^{\rm max} < 1$
is observed. The smaller the $l$-doubling, the larger is the $|P|^{\rm max},$ \cite{petrov2022sensitivity}.
Thus, the value of $l$-doubling directly influences the sensitivity of linear triatomic molecules to the $\mathcal{P}$,$\mathcal{T}$-odd effects.

However, when calculating $l$-doubling, in our preceding work \cite{zakharova2021rovibrational} we assumed that the deformations of the ligand may be neglected as the corresponding vibrational frequency is much higher than the frequencies for bending and stretching of the whole molecule. The aim of the present paper is to take into account the change of the ligand's bond length
(i.e., of the equilibrium ligand size and its vibrational frequency) due to the interaction with bending and stretching modes of the YbOH.

\section{Potential surface interpolation}

To obtain the potential surface we used the Dirac 19 program suite \cite{DIRAC19}. The PNPI \cite{QCPNPI:Basis} 42-valence electron basis was used to describe the Ytterbium atom. Moreover, to simplify calculations with a heavy atom, we employed the 28-electron generalized relativistic effective potential, also developed by the PNPI Quantum Chemistry Laboratory \cite{titov1999generalized,mosyagin2010shape,mosyagin2016generalized}. For oxygen and hydrogen atoms, we used cc-pVTZ basis sets. Active space for the coupled cluster computations comprises 21 active and 30 frozen electrons.

\begin{figure}[h]
\centering
  \includegraphics[width=0.45\textwidth]{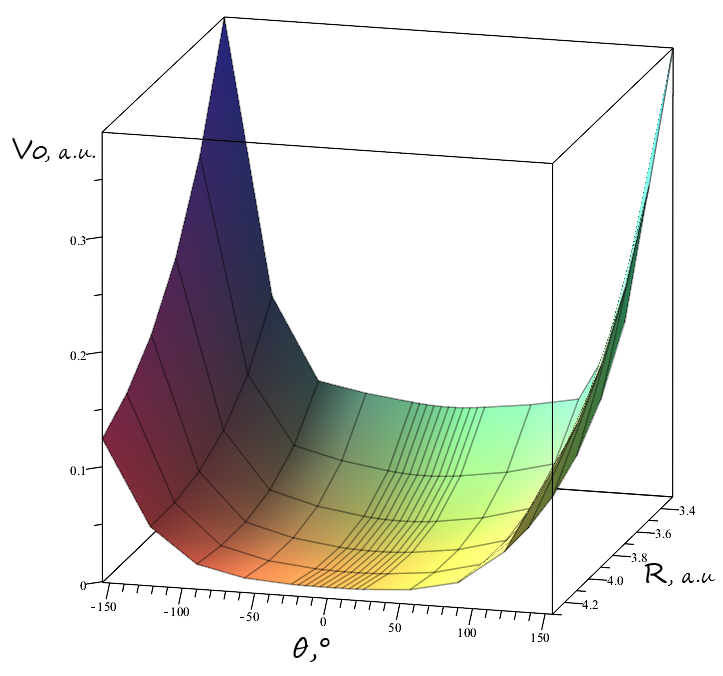}
  \caption{Morse potential parameter $V_0$ as a function of $R$ and $\theta$ coordinates}
  \label{V0Plot}
\end{figure}

\begin{figure}[h]
\centering
  \includegraphics[width=0.45\textwidth]{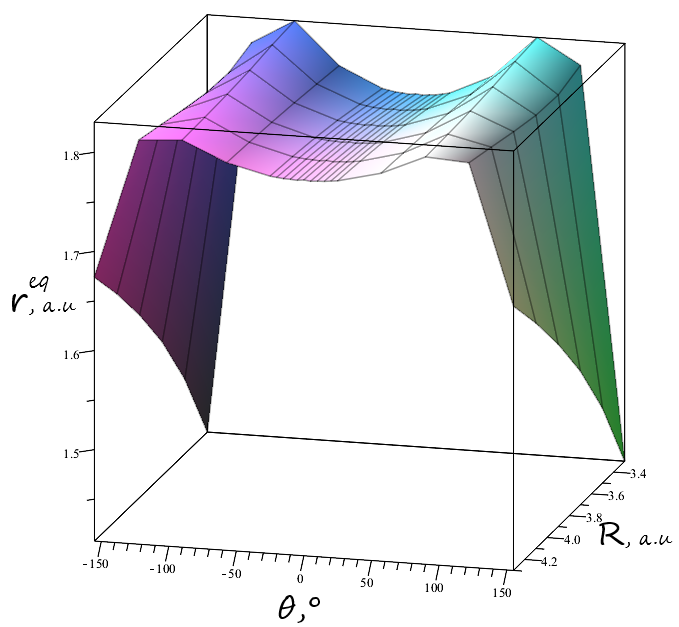}
  \caption{Morse potential parameter $r^{eq}$ as a function of $R$ and $\theta$ coordinates}
  \label{ReqPlot}
\end{figure}

We obtain the CCSD and CCSD(T) potential surface $V$ on the grid of coordinates $(R_i,r_j,\theta_k)$,
\begin{align}
\{R_i\}=3.3, 3.5,\ldots 4.3\,\mathrm{a.u.}\\
\{r_j\}=1.632, 1.732,\ldots 2.032\,\mathrm{a.u.}\\
\{\theta_k\}=0^\circ, 5^\circ, 10^\circ, 15^\circ, 20^\circ, 25^\circ, 55^\circ, 90^\circ, 122^\circ, 155^\circ
\end{align}
At the first step we approximate the dependence on $r$ for each $(R_i,\theta_k)$ by the Morse potential
\begin{equation}
    V(r) = V_0+D(1-e^{-\alpha(r-r^{eq})})^2
\end{equation}
where $r^{eq}$ is the equilibrium ligand size, $D$ -- dissociation energy, $\alpha$ is the exponent parameter, and $V_0$ is the energy of the minimum. The values for $V_0$ and $r^{eq}$ parameters for various configurations of YbOH are presented on Fig. \ref{V0Plot} and Fig. \ref{ReqPlot} correspondingly.

At the second step, for each $R_i$ we interpolate the dependence of this set of parameters $V_0(R_i,
\theta_k)$, $r^{eq}(R_i,\theta_k)$, $D(R_i,
\theta_k)$, and $
\alpha(R_i,
\theta_k)$, on $\theta$ by the Akima splines. 

Then, using Akima splines, obtained at the previous step, we take the values at the zeros of the $(\lambda_{max}+1)$-th Legendre polynomial $P_{\lambda_{max}+1}(x)$, where $x=\cos\theta$. For each $R_i$, we then use the Gauss-Legendre method to represent the parameters as sums of Legendre polynomials,
\begin{align}
V_0(R_i,\theta)=\sum_{\lambda=0}^{\lambda_{max}}V_{0,\lambda}(R_i)P_\lambda(
\cos\theta),\\
D(R_i,\theta)=\sum_{\lambda=0}^{\lambda_{max}}D_{\lambda}(R_i)P_\lambda(
\cos\theta),\\
r^{eq}(R_i,\theta)=\sum_{\lambda=0}^{\lambda_{max}}r^{eq}_{\lambda}(R_i)P_\lambda(
\cos\theta),\\
\alpha(R_i,\theta)=\sum_{\lambda=0}^{\lambda_{max}}\alpha_{\lambda}(R_i)P_\lambda(
\cos\theta),
\end{align}
In this paper we have taken $\lambda_{max}=40$. Such a representation is guaranteed to be smooth and have good periodicity in $\theta$.

At the final step, we interpolate the dependence of the coefficients of the Legendre polynomials on $R$ using Akima's splines.


\section{Methods}

We use the Born-Oppenheimer approximation to separate wavefunction to $\Psi_{\rm nuc}$,
responsible for the nuclear motion in the adiabatic potential of the electrons, and the $\Psi_{\rm elec}$, corresponding to the electronic motion in the field of heavy nuclei
\begin{equation}
\Psi_{\rm total}\approx\Psi_{\rm nuc}(\vec{R}, \vec{r})\psi_{\rm elec}(\vec{R}, \vec{r}|q),
\label{totalWF}
\end{equation}

where $q$ are the generalized coordinates of the electrons, $\vec{R}$ is the vector from the Ytterbium atom to the center of mass of the OH ligand, $\vec{r}$ is the vector from the Oxygen to the Hydrogen atom.

\begin{figure}[h]
\centering
  \includegraphics[width=0.25\textwidth]{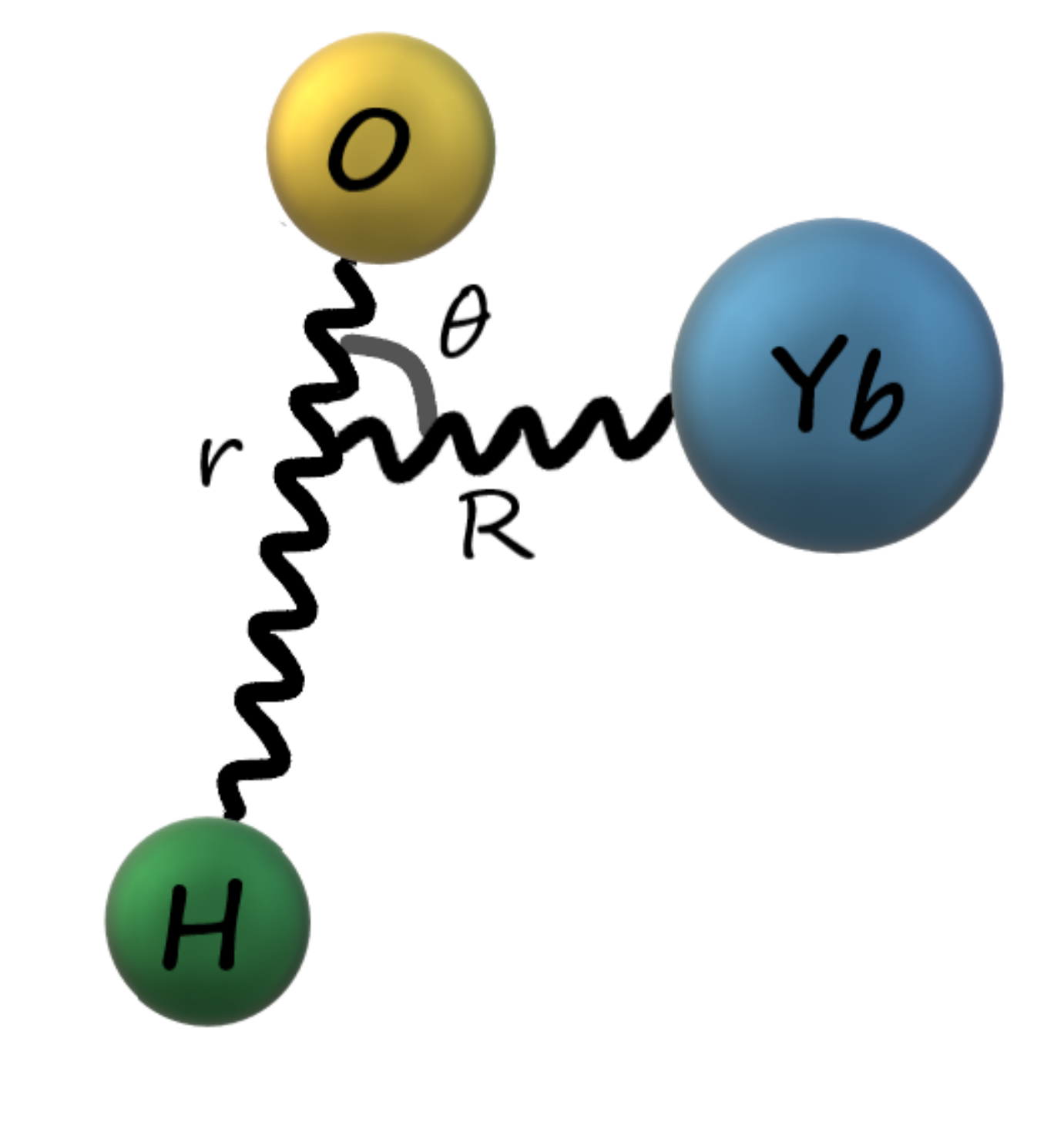}
  \caption{Jacobi coordinates}
  \label{Jacob}
\end{figure}

The 
Hamiltonian of nuclei takes the form,
\begin{equation}
\hat{H}_{\rm nuc}=-\frac{1}{2\mu}\frac{\partial^2}{\partial R^2} -\frac{1}{2\mu_{OH}}\frac{\partial^2}{\partial r^2}+\frac{\hat{L}^2}{2\mu R^2}+\frac{\hat{j}^2}{2\mu_{OH}r^2}+V(R, r,\theta),
\end{equation}
where $\theta$ is the angle between vectors $\vec{R}$ and $\vec{r}$ (so that $\theta=0$ corresponds to the Yb-O-H linear configuration), $\mu$ is the $Yb-OH$ reduced mass, $\mu_{OH}$ is the  reduced mass of the ligand,
$\hat{L}$ and $\hat{j}$  are the angular momentum of the whole system rotation and the OH angular momentum correspondingly. Directions of axes $\hat{r}$ and $\hat{R}$  are represented in Fig.~\ref{Jacob}. $V$ is the electronic potential surface  obtained on the CCSD or CCSD(T) level and depending only on the relative Jacobi coordinates $(R, r, \theta)$.

The nuclear wavefunction $\Psi_{\rm nuc}(\vec{R}, \vec{r})$ is the solution of the Schr\"{o}dinger equation
\begin{equation}
\hat{H}_{nuc}\Psi_{\rm nuc}(\vec{R}, \vec{r}) = E \Psi_{\rm nuc}(\vec{R}, \vec{r}).
\label{Shreq}
\end{equation}
To solve eq. (\ref{Shreq}) we use the expansion
\begin{equation}
\Psi_{\rm nuc}(\vec{R}, \vec{r}) = \sum_{L=0}^{L_{max}}\sum_{j=0}^{j_{max}}\sum_{n=1}^{n_{max}} F_{JjLn}(R)\Phi_{JjLM}(\hat{R},\hat{r})f_n(r),
\label{psiexp}
\end{equation}
where
\begin{equation}
\Phi_{JjLM}(\hat{R},\hat{r}) = \sum_{m_L,m_j} C^{JM}_{Lm_L,jm_j} Y_{Lm_L}(\hat{R})Y_{jm_j}(\hat{r})
\end{equation}
is coupled to conserved total angular momentum $J$ basis set, $Y_{Lm_L}$ is a spherical function, $f_n(r)$ is a solution of
\begin{equation}
 \left(-\frac{1}{2\mu_{OH}}\frac{\partial^2}{\partial r^2}+V(R_i, r,\theta_i)\right)f_n(r) = e_n f_n(r),
 \label{wOH}
\end{equation}
where $R_i$ and $\theta_i$ are some fixed values. To test the approximation (\ref{psiexp}) we perform our computations for different sets of $(R_i,\theta_i)$.

Substituting  wavefunction (\ref{psiexp}) to eq. (\ref{Shreq}) one gets
the system of close-coupled equations for $F_{JjLn}(R)$ \cite{mcguire1974quantum}.

\section{Results}

The parameters obtained from the spectrum of the nuclear wavefunctions $\Psi_{\rm nuc}$ compared with the results from the rigid ligand approximation and experimental values are presented in the Table~\ref{tab:Spectrum}. One can see that taking into account the ligand deformation improves the agreement with experiments. To test our method and codes, which now take into account the ligand deformation, we performed calculations
with two different sets of the parameters $R_i$ and $\theta_i$ from eq.(\ref{wOH}).
The Set 1 corresponds to $R_1=3.9$ a.u, $\theta_1=0$ and 
and Set 2 corresponds to $R_2=4.2$ a.u, $\theta_1=0.5$ radians.
In Table~\ref{tab:Conv} one can see that results with Set 1 and Set 2 converge to each other as quantum number $n_{max}$ (see eq. (\ref{psiexp})) increases. 
Table~\ref{tab:Conv} also show that for purposes of the present paper it is sufficient to use $n_{max}=2$.
Probably inclusion of the noniterative triple excitations to the coupled cluster computations and using more extensive basis sets are necessary to reach the accuracy of the calculation on the order of $
\sim 5
\,\mathrm{cm}^{-1}$ for the vibrational frequencies, what is out of scope of the current work.

However, to estimate the actual value of the $l$-doubling one can note the following.
The decrease of the $l$-doubling value together with the increase of the $\nu_2$ frequency is consistent with the estimate \cite{HerzbergBook},
\begin{equation}
q \simeq 
    \frac{B^2}{\nu_2}\Big(1+4\sum_{k=1,3}\frac{\zeta_{2k}^2\nu_{2}^2}{\nu_{k}^2-\nu_{2}^2}\Big)(v+1),\quad \Delta E=2q,
\end{equation}
where $\zeta_{2k}$ are Coriolis coefficients.
As the current experimental data for the $\nu_2$ frequency fall into a rather wide range $319-339
,\mathrm{cm}^{-1}$, we sum up our results for $l$-doubling with the estimate of $\Delta E_{J=1}
\simeq 24-26\,\mathrm{MHz}$.

\begin{table*}
\caption{\label{tab:Spectrum} Rovibrational spectrum parameters}
\begin{ruledtabular}
\begin{tabular}{ccccc}
Parameter& Rigid ligand, CCSD(T)\cite{zakharova2021rovibrational}  & CCSD & CCSD(T) & Experiment\footnote{The number in parentheses denotes $2\sigma$ deviation} \\
\hline
Stretching mode $\nu_1$, ${\rm cm}^{-1}$
  & 550    & 545   & 536  & 529.341(1)\cite{melville2001visible}\\
Bending mode $\nu_2$, ${\rm cm}^{-1}$
  & 319   & 351  & 342  & 319(5)\cite{zhang2021accurate}, 329(5) \cite{mengesha2020branching}, 339(5),\cite{melville2001visible} \\
Ligand mode $\nu_3$, ${\rm cm}^{-1}$ & -  & 4055  & 4030 & -\\
Rotational constant $B(\nu_1=0,\nu_2=0,\nu_3=0)$, ${\rm cm}^{-1}$  & 0.2461   & 0.2456  & 0.2468 & 0.245434(13) \cite{melville2001visible}\\

Rotational constant $B(\nu_1=1,\nu_2=0,\nu_3=0)$, ${\rm cm}^{-1}$  & -  & 0.2437  & 0.2443 & 0.2439771(35)\cite{melville2001visible}\\

$l$-doubling $\Delta E_{J=1}=2q$, MHz  & 26   & 23 & 24 & -
\end{tabular}
\end{ruledtabular}
\end{table*}

\begin{table*}
\caption{\label{tab:Conv} Convergence of the results with $n_{max}$ (see eq. (\ref{psiexp})) for different sets of parameters $R_i$ and $\theta_i$ (see eq.(\ref{wOH})) for CCSD(T) calculations}
\begin{ruledtabular}
\begin{tabular}{ccccc}
parameter                     & Set of $R_i,\theta_i$  &   $n_{max}=1$ & $n_{max}=2$ & $n_{max}=3$ \\
Bending mode $\nu_2$, cm$^{-1}$ &     1           &  344.0892     &  341.9777   &  341.9665   \\
                              &     2             &  340.9922    &  341.9698   &  341.9671 \\
$l$-doubling $\Delta E_{J=1}=2q$, MHz &     1     &  23.520     &   24.247    &  24.257    \\
                              &     2             &  23.712     &   24.235    &  24.256
\end{tabular}
\end{ruledtabular}
\end{table*}

\begin{acknowledgments}
The work is supported by the Russian Science Foundation grant No. 18-12-00227.
\end{acknowledgments}
\section*{Author Declarations}

\subsection*{Conflict of interest}
The authors have no conflicts to disclose.

\section*{Availability of data}
The data that support the findings of this study are available from the corresponding author upon reasonable request.


\begin{thebibliography}{49}%
\makeatletter
\providecommand \@ifxundefined [1]{%
 \@ifx{#1\undefined}
}%
\providecommand \@ifnum [1]{%
 \ifnum #1\expandafter \@firstoftwo
 \else \expandafter \@secondoftwo
 \fi
}%
\providecommand \@ifx [1]{%
 \ifx #1\expandafter \@firstoftwo
 \else \expandafter \@secondoftwo
 \fi
}%
\providecommand \natexlab [1]{#1}%
\providecommand \enquote  [1]{``#1''}%
\providecommand \bibnamefont  [1]{#1}%
\providecommand \bibfnamefont [1]{#1}%
\providecommand \citenamefont [1]{#1}%
\providecommand \href@noop [0]{\@secondoftwo}%
\providecommand \href [0]{\begingroup \@sanitize@url \@href}%
\providecommand \@href[1]{\@@startlink{#1}\@@href}%
\providecommand \@@href[1]{\endgroup#1\@@endlink}%
\providecommand \@sanitize@url [0]{\catcode `\\12\catcode `\$12\catcode
  `\&12\catcode `\#12\catcode `\^12\catcode `\_12\catcode `\%12\relax}%
\providecommand \@@startlink[1]{}%
\providecommand \@@endlink[0]{}%
\providecommand \url  [0]{\begingroup\@sanitize@url \@url }%
\providecommand \@url [1]{\endgroup\@href {#1}{\urlprefix }}%
\providecommand \urlprefix  [0]{URL }%
\providecommand \Eprint [0]{\href }%
\providecommand \doibase [0]{https://doi.org/}%
\providecommand \selectlanguage [0]{\@gobble}%
\providecommand \bibinfo  [0]{\@secondoftwo}%
\providecommand \bibfield  [0]{\@secondoftwo}%
\providecommand \translation [1]{[#1]}%
\providecommand \BibitemOpen [0]{}%
\providecommand \bibitemStop [0]{}%
\providecommand \bibitemNoStop [0]{.\EOS\space}%
\providecommand \EOS [0]{\spacefactor3000\relax}%
\providecommand \BibitemShut  [1]{\csname bibitem#1\endcsname}%
\let\auto@bib@innerbib\@empty
\bibitem [{\citenamefont {Petrov}\ and\ \citenamefont
  {Zakharova}(2022)}]{petrov2022sensitivity}%
  \BibitemOpen
  \bibfield  {author} {\bibinfo {author} {\bibfnamefont {A.}~\bibnamefont
  {Petrov}}\ and\ \bibinfo {author} {\bibfnamefont {A.}~\bibnamefont
  {Zakharova}},\ }\bibfield  {title} {\enquote {\bibinfo {title} {Sensitivity
  of the yboh molecule to p t-odd effects in an external electric field},}\
  }\href@noop {} {\bibfield  {journal} {\bibinfo  {journal} {Physical Review
  A}\ }\textbf {\bibinfo {volume} {105}},\ \bibinfo {pages} {L050801} (\bibinfo
  {year} {2022})}\BibitemShut {NoStop}%
\bibitem [{\citenamefont {Zakharova}, \citenamefont {Kurchavov},\ and\
  \citenamefont {Petrov}(2021)}]{zakharova2021rovibrational}%
  \BibitemOpen
  \bibfield  {author} {\bibinfo {author} {\bibfnamefont {A.}~\bibnamefont
  {Zakharova}}, \bibinfo {author} {\bibfnamefont {I.}~\bibnamefont
  {Kurchavov}},\ and\ \bibinfo {author} {\bibfnamefont {A.}~\bibnamefont
  {Petrov}},\ }\bibfield  {title} {\enquote {\bibinfo {title} {Rovibrational
  structure of the ytterbium monohydroxide molecule and the p, t-violation
  searches},}\ }\href@noop {} {\bibfield  {journal} {\bibinfo  {journal} {The
  Journal of Chemical Physics}\ }\textbf {\bibinfo {volume} {155}},\ \bibinfo
  {pages} {164301} (\bibinfo {year} {2021})}\BibitemShut {NoStop}%
\bibitem [{\citenamefont {Schwartz}(2014)}]{schwartz2014quantum}%
  \BibitemOpen
  \bibfield  {author} {\bibinfo {author} {\bibfnamefont {M.~D.}\ \bibnamefont
  {Schwartz}},\ }\href@noop {} {\emph {\bibinfo {title} {Quantum field theory
  and the standard model}}}\ (\bibinfo  {publisher} {Cambridge University
  Press},\ \bibinfo {year} {2014})\BibitemShut {NoStop}%
\bibitem [{\citenamefont {{Particle Data Group}}\ \emph
  {et~al.}(2020)\citenamefont {{Particle Data Group}}, \citenamefont {Zyla},
  \citenamefont {Barnett}, \citenamefont {Beringer}, \citenamefont {Dahl},
  \citenamefont {Dwyer}, \citenamefont {Groom}, \citenamefont {Lin},
  \citenamefont {Lugovsky}, \citenamefont {Pianori} \emph
  {et~al.}}]{particle2020review}%
  \BibitemOpen
  \bibfield  {author} {\bibinfo {author} {\bibnamefont {{Particle Data
  Group}}}, \bibinfo {author} {\bibfnamefont {P.}~\bibnamefont {Zyla}},
  \bibinfo {author} {\bibfnamefont {R.}~\bibnamefont {Barnett}}, \bibinfo
  {author} {\bibfnamefont {J.}~\bibnamefont {Beringer}}, \bibinfo {author}
  {\bibfnamefont {O.}~\bibnamefont {Dahl}}, \bibinfo {author} {\bibfnamefont
  {D.}~\bibnamefont {Dwyer}}, \bibinfo {author} {\bibfnamefont
  {D.}~\bibnamefont {Groom}}, \bibinfo {author} {\bibfnamefont {C.-J.}\
  \bibnamefont {Lin}}, \bibinfo {author} {\bibfnamefont {K.}~\bibnamefont
  {Lugovsky}}, \bibinfo {author} {\bibfnamefont {E.}~\bibnamefont {Pianori}},
  \emph {et~al.},\ }\bibfield  {title} {\enquote {\bibinfo {title} {Review of
  particle physics},}\ }\href@noop {} {\bibfield  {journal} {\bibinfo
  {journal} {Progress of Theoretical and Experimental Physics}\ }\textbf
  {\bibinfo {volume} {2020}},\ \bibinfo {pages} {083C01} (\bibinfo {year}
  {2020})}\BibitemShut {NoStop}%
\bibitem [{\citenamefont {Cheng}(1988)}]{Cheng1988}%
  \BibitemOpen
  \bibfield  {author} {\bibinfo {author} {\bibfnamefont {H.-Y.}\ \bibnamefont
  {Cheng}},\ }\bibfield  {title} {\enquote {\bibinfo {title} {The strong {CP}
  problem revisited},}\ }\href@noop {} {\bibfield  {journal} {\bibinfo
  {journal} {Physics Reports}\ }\textbf {\bibinfo {volume} {158}},\ \bibinfo
  {pages} {1--89} (\bibinfo {year} {1988})}\BibitemShut {NoStop}%
\bibitem [{\citenamefont {Kim}\ and\ \citenamefont
  {Carosi}(2010)}]{KimCarosi2010}%
  \BibitemOpen
  \bibfield  {author} {\bibinfo {author} {\bibfnamefont {J.~E.}\ \bibnamefont
  {Kim}}\ and\ \bibinfo {author} {\bibfnamefont {G.}~\bibnamefont {Carosi}},\
  }\bibfield  {title} {\enquote {\bibinfo {title} {{Axions and the Strong CP
  Problem}},}\ }\href {https://doi.org/10.1103/RevModPhys.82.557} {\bibfield
  {journal} {\bibinfo  {journal} {Rev. Mod. Phys.}\ }\textbf {\bibinfo {volume}
  {82}},\ \bibinfo {pages} {557--602} (\bibinfo {year} {2010})},\ \bibinfo
  {note} {[Erratum: Rev.Mod.Phys. 91, 049902 (2019)]},\ \Eprint
  {https://arxiv.org/abs/0807.3125} {arXiv:0807.3125 [hep-ph]} \BibitemShut
  {NoStop}%
\bibitem [{\citenamefont {Khriplovich}\ and\ \citenamefont
  {Lamoreaux}(2012)}]{khriplovich2012cp}%
  \BibitemOpen
  \bibfield  {author} {\bibinfo {author} {\bibfnamefont {I.~B.}\ \bibnamefont
  {Khriplovich}}\ and\ \bibinfo {author} {\bibfnamefont {S.~K.}\ \bibnamefont
  {Lamoreaux}},\ }\href@noop {} {\emph {\bibinfo {title} {CP violation without
  strangeness: electric dipole moments of particles, atoms, and molecules}}}\
  (\bibinfo  {publisher} {Springer Science \& Business Media},\ \bibinfo {year}
  {2012})\BibitemShut {NoStop}%
\bibitem [{\citenamefont {Cabibbo}(1963)}]{Cabibbo1963}%
  \BibitemOpen
  \bibfield  {author} {\bibinfo {author} {\bibfnamefont {N.}~\bibnamefont
  {Cabibbo}},\ }\bibfield  {title} {\enquote {\bibinfo {title} {{Unitary
  Symmetry and Leptonic Decays}},}\ }\href
  {https://doi.org/10.1103/PhysRevLett.10.531} {\bibfield  {journal} {\bibinfo
  {journal} {Phys. Rev. Lett.}\ }\textbf {\bibinfo {volume} {10}},\ \bibinfo
  {pages} {531--533} (\bibinfo {year} {1963})}\BibitemShut {NoStop}%
\bibitem [{\citenamefont {Kobayashi}\ and\ \citenamefont
  {Maskawa}(1973)}]{KobayashiMaskawa1973}%
  \BibitemOpen
  \bibfield  {author} {\bibinfo {author} {\bibfnamefont {M.}~\bibnamefont
  {Kobayashi}}\ and\ \bibinfo {author} {\bibfnamefont {T.}~\bibnamefont
  {Maskawa}},\ }\bibfield  {title} {\enquote {\bibinfo {title} {{CP Violation
  in the Renormalizable Theory of Weak Interaction}},}\ }\href
  {https://doi.org/10.1143/PTP.49.652} {\bibfield  {journal} {\bibinfo
  {journal} {Prog. Theor. Phys.}\ }\textbf {\bibinfo {volume} {49}},\ \bibinfo
  {pages} {652--657} (\bibinfo {year} {1973})}\BibitemShut {NoStop}%
\bibitem [{\citenamefont {Pontecorvo}(1957)}]{Pontecorvo1957}%
  \BibitemOpen
  \bibfield  {author} {\bibinfo {author} {\bibfnamefont {B.}~\bibnamefont
  {Pontecorvo}},\ }\bibfield  {title} {\enquote {\bibinfo {title} {{Inverse
  beta processes and nonconservation of lepton charge}},}\ }\href@noop {}
  {\bibfield  {journal} {\bibinfo  {journal} {Zh. Eksp. Teor. Fiz.}\ }\textbf
  {\bibinfo {volume} {34}},\ \bibinfo {pages} {247} (\bibinfo {year}
  {1957})}\BibitemShut {NoStop}%
\bibitem [{\citenamefont {Maki}, \citenamefont {Nakagawa},\ and\ \citenamefont
  {Sakata}(1962)}]{MNS1962}%
  \BibitemOpen
  \bibfield  {author} {\bibinfo {author} {\bibfnamefont {Z.}~\bibnamefont
  {Maki}}, \bibinfo {author} {\bibfnamefont {M.}~\bibnamefont {Nakagawa}},\
  and\ \bibinfo {author} {\bibfnamefont {S.}~\bibnamefont {Sakata}},\
  }\bibfield  {title} {\enquote {\bibinfo {title} {{Remarks on the unified
  model of elementary particles}},}\ }\href
  {https://doi.org/10.1143/PTP.28.870} {\bibfield  {journal} {\bibinfo
  {journal} {Prog. Theor. Phys.}\ }\textbf {\bibinfo {volume} {28}},\ \bibinfo
  {pages} {870--880} (\bibinfo {year} {1962})}\BibitemShut {NoStop}%
\bibitem [{\citenamefont {Barr}(1992)}]{barr1992t}%
  \BibitemOpen
  \bibfield  {author} {\bibinfo {author} {\bibfnamefont {S.~M.}\ \bibnamefont
  {Barr}},\ }\bibfield  {title} {\enquote {\bibinfo {title} {T-and p-odd
  electron-nucleon interactions and the electric dipole moments of large
  atoms},}\ }\href@noop {} {\bibfield  {journal} {\bibinfo  {journal} {Physical
  Review D}\ }\textbf {\bibinfo {volume} {45}},\ \bibinfo {pages} {4148}
  (\bibinfo {year} {1992})}\BibitemShut {NoStop}%
\bibitem [{\citenamefont {FUKUYAMA}(2012)}]{Fukuyama2012}%
  \BibitemOpen
  \bibfield  {author} {\bibinfo {author} {\bibfnamefont {T.}~\bibnamefont
  {FUKUYAMA}},\ }\bibfield  {title} {\enquote {\bibinfo {title} {{SEARCHING FOR
  NEW PHYSICS BEYOND THE STANDARD MODEL IN ELECTRIC DIPOLE MOMENT}},}\ }\href
  {https://doi.org/10.1142/S0217751X12300153} {\bibfield  {journal} {\bibinfo
  {journal} {International Journal of Modern Physics A}\ }\textbf {\bibinfo
  {volume} {27}},\ \bibinfo {pages} {1230015} (\bibinfo {year}
  {2012})}\BibitemShut {NoStop}%
\bibitem [{\citenamefont {Pospelov}\ and\ \citenamefont
  {Ritz}(2014)}]{PospelovRitz2014}%
  \BibitemOpen
  \bibfield  {author} {\bibinfo {author} {\bibfnamefont {M.}~\bibnamefont
  {Pospelov}}\ and\ \bibinfo {author} {\bibfnamefont {A.}~\bibnamefont
  {Ritz}},\ }\bibfield  {title} {\enquote {\bibinfo {title} {{CKM benchmarks
  for electron electric dipole moment experiments}},}\ }\href
  {https://doi.org/10.1103/PhysRevD.89.056006} {\bibfield  {journal} {\bibinfo
  {journal} {Phys. Rev. D}\ }\textbf {\bibinfo {volume} {89}},\ \bibinfo
  {pages} {056006} (\bibinfo {year} {2014})},\ \Eprint
  {https://arxiv.org/abs/1311.5537} {arXiv:1311.5537 [hep-ph]} \BibitemShut
  {NoStop}%
\bibitem [{\citenamefont {Yamaguchi}\ and\ \citenamefont
  {Yamanaka}(2020)}]{YamaguchiYamanaka2020}%
  \BibitemOpen
  \bibfield  {author} {\bibinfo {author} {\bibfnamefont {Y.}~\bibnamefont
  {Yamaguchi}}\ and\ \bibinfo {author} {\bibfnamefont {N.}~\bibnamefont
  {Yamanaka}},\ }\bibfield  {title} {\enquote {\bibinfo {title} {{Large
  long-distance contributions to the electric dipole moments of charged leptons
  in the standard model}},}\ }\href
  {https://doi.org/10.1103/PhysRevLett.125.241802} {\bibfield  {journal}
  {\bibinfo  {journal} {Phys. Rev. Lett.}\ }\textbf {\bibinfo {volume} {125}},\
  \bibinfo {pages} {241802} (\bibinfo {year} {2020})},\ \Eprint
  {https://arxiv.org/abs/2003.08195} {arXiv:2003.08195 [hep-ph]} \BibitemShut
  {NoStop}%
\bibitem [{\citenamefont {Yamaguchi}\ and\ \citenamefont
  {Yamanaka}(2021)}]{YamaguchiYamanaka2021}%
  \BibitemOpen
  \bibfield  {author} {\bibinfo {author} {\bibfnamefont {Y.}~\bibnamefont
  {Yamaguchi}}\ and\ \bibinfo {author} {\bibfnamefont {N.}~\bibnamefont
  {Yamanaka}},\ }\bibfield  {title} {\enquote {\bibinfo {title} {{Quark level
  and hadronic contributions to the electric dipole moment of charged leptons
  in the standard model}},}\ }\href
  {https://doi.org/10.1103/PhysRevD.103.013001} {\bibfield  {journal} {\bibinfo
   {journal} {Phys. Rev. D}\ }\textbf {\bibinfo {volume} {103}},\ \bibinfo
  {pages} {013001} (\bibinfo {year} {2021})},\ \Eprint
  {https://arxiv.org/abs/2006.00281} {arXiv:2006.00281 [hep-ph]} \BibitemShut
  {NoStop}%
\bibitem [{\citenamefont {Baron}\ \emph {et~al.}(2014)\citenamefont {Baron},
  \citenamefont {Campbell}, \citenamefont {DeMille}, \citenamefont {Doyle},
  \citenamefont {Gabrielse}, \citenamefont {Gurevich}, \citenamefont {Hess},
  \citenamefont {Hutzler}, \citenamefont {Kirilov}, \citenamefont {Kozyryev}
  \emph {et~al.}}]{baron2014order}%
  \BibitemOpen
  \bibfield  {author} {\bibinfo {author} {\bibfnamefont {J.}~\bibnamefont
  {Baron}}, \bibinfo {author} {\bibfnamefont {W.~C.}\ \bibnamefont {Campbell}},
  \bibinfo {author} {\bibfnamefont {D.}~\bibnamefont {DeMille}}, \bibinfo
  {author} {\bibfnamefont {J.~M.}\ \bibnamefont {Doyle}}, \bibinfo {author}
  {\bibfnamefont {G.}~\bibnamefont {Gabrielse}}, \bibinfo {author}
  {\bibfnamefont {Y.~V.}\ \bibnamefont {Gurevich}}, \bibinfo {author}
  {\bibfnamefont {P.~W.}\ \bibnamefont {Hess}}, \bibinfo {author}
  {\bibfnamefont {N.~R.}\ \bibnamefont {Hutzler}}, \bibinfo {author}
  {\bibfnamefont {E.}~\bibnamefont {Kirilov}}, \bibinfo {author} {\bibfnamefont
  {I.}~\bibnamefont {Kozyryev}}, \emph {et~al.},\ }\bibfield  {title} {\enquote
  {\bibinfo {title} {Order of magnitude smaller limit on the electric dipole
  moment of the electron},}\ }\href@noop {} {\bibfield  {journal} {\bibinfo
  {journal} {Science}\ }\textbf {\bibinfo {volume} {343}},\ \bibinfo {pages}
  {269--272} (\bibinfo {year} {2014})}\BibitemShut {NoStop}%
\bibitem [{\citenamefont {DeMille}, \citenamefont {Doyle},\ and\ \citenamefont
  {Sushkov}(2017)}]{demille2017probing}%
  \BibitemOpen
  \bibfield  {author} {\bibinfo {author} {\bibfnamefont {D.}~\bibnamefont
  {DeMille}}, \bibinfo {author} {\bibfnamefont {J.~M.}\ \bibnamefont {Doyle}},\
  and\ \bibinfo {author} {\bibfnamefont {A.~O.}\ \bibnamefont {Sushkov}},\
  }\bibfield  {title} {\enquote {\bibinfo {title} {Probing the frontiers of
  particle physics with tabletop-scale experiments},}\ }\href@noop {}
  {\bibfield  {journal} {\bibinfo  {journal} {Science}\ }\textbf {\bibinfo
  {volume} {357}},\ \bibinfo {pages} {990--994} (\bibinfo {year}
  {2017})}\BibitemShut {NoStop}%
\bibitem [{\citenamefont {Andreev}\ \emph {et~al.}(2018)\citenamefont
  {Andreev}, \citenamefont {Ang}, \citenamefont {DeMille}, \citenamefont
  {Doyle}, \citenamefont {Gabrielse}, \citenamefont {Haefner}, \citenamefont
  {Hutzler}, \citenamefont {Lasner}, \citenamefont {Meisenhelder},
  \citenamefont {O'Leary} \emph {et~al.}}]{ACME:18}%
  \BibitemOpen
  \bibfield  {author} {\bibinfo {author} {\bibfnamefont {V.}~\bibnamefont
  {Andreev}}, \bibinfo {author} {\bibfnamefont {D.}~\bibnamefont {Ang}},
  \bibinfo {author} {\bibfnamefont {D.}~\bibnamefont {DeMille}}, \bibinfo
  {author} {\bibfnamefont {J.}~\bibnamefont {Doyle}}, \bibinfo {author}
  {\bibfnamefont {G.}~\bibnamefont {Gabrielse}}, \bibinfo {author}
  {\bibfnamefont {J.}~\bibnamefont {Haefner}}, \bibinfo {author} {\bibfnamefont
  {N.}~\bibnamefont {Hutzler}}, \bibinfo {author} {\bibfnamefont
  {Z.}~\bibnamefont {Lasner}}, \bibinfo {author} {\bibfnamefont
  {C.}~\bibnamefont {Meisenhelder}}, \bibinfo {author} {\bibfnamefont
  {B.}~\bibnamefont {O'Leary}}, \emph {et~al.},\ }\bibfield  {title} {\enquote
  {\bibinfo {title} {Improved limit on the electric dipole moment of the
  electron},}\ }\href@noop {} {\bibfield  {journal} {\bibinfo  {journal}
  {Nature}\ }\textbf {\bibinfo {volume} {562}},\ \bibinfo {pages} {355--360}
  (\bibinfo {year} {2018})}\BibitemShut {NoStop}%
\bibitem [{\citenamefont {Cirigliano}\ \emph {et~al.}(2019)\citenamefont
  {Cirigliano}, \citenamefont {Crivellin}, \citenamefont {Dekens},
  \citenamefont {de~Vries}, \citenamefont {Hoferichter},\ and\ \citenamefont
  {Mereghetti}}]{cirigliano2019c}%
  \BibitemOpen
  \bibfield  {author} {\bibinfo {author} {\bibfnamefont {V.}~\bibnamefont
  {Cirigliano}}, \bibinfo {author} {\bibfnamefont {A.}~\bibnamefont
  {Crivellin}}, \bibinfo {author} {\bibfnamefont {W.}~\bibnamefont {Dekens}},
  \bibinfo {author} {\bibfnamefont {J.}~\bibnamefont {de~Vries}}, \bibinfo
  {author} {\bibfnamefont {M.}~\bibnamefont {Hoferichter}},\ and\ \bibinfo
  {author} {\bibfnamefont {E.}~\bibnamefont {Mereghetti}},\ }\bibfield  {title}
  {\enquote {\bibinfo {title} {C p violation in higgs-gauge interactions: From
  tabletop experiments to the lhc},}\ }\href@noop {} {\bibfield  {journal}
  {\bibinfo  {journal} {Physical Review Letters}\ }\textbf {\bibinfo {volume}
  {123}},\ \bibinfo {pages} {051801} (\bibinfo {year} {2019})}\BibitemShut
  {NoStop}%
\bibitem [{\citenamefont {Hutzler}\ \emph {et~al.}(2020)\citenamefont
  {Hutzler}, \citenamefont {Borschevsky}, \citenamefont {Budker}, \citenamefont
  {DeMille}, \citenamefont {Flambaum}, \citenamefont {Gabrielse}, \citenamefont
  {Ruiz}, \citenamefont {Jayich}, \citenamefont {Orozco}, \citenamefont
  {Ramsey-Musolf} \emph {et~al.}}]{hutzler2020searches}%
  \BibitemOpen
  \bibfield  {author} {\bibinfo {author} {\bibfnamefont {N.~R.}\ \bibnamefont
  {Hutzler}}, \bibinfo {author} {\bibfnamefont {A.}~\bibnamefont
  {Borschevsky}}, \bibinfo {author} {\bibfnamefont {D.}~\bibnamefont {Budker}},
  \bibinfo {author} {\bibfnamefont {D.}~\bibnamefont {DeMille}}, \bibinfo
  {author} {\bibfnamefont {V.}~\bibnamefont {Flambaum}}, \bibinfo {author}
  {\bibfnamefont {G.}~\bibnamefont {Gabrielse}}, \bibinfo {author}
  {\bibfnamefont {R.}~\bibnamefont {Ruiz}}, \bibinfo {author} {\bibfnamefont
  {A.}~\bibnamefont {Jayich}}, \bibinfo {author} {\bibfnamefont
  {L.}~\bibnamefont {Orozco}}, \bibinfo {author} {\bibfnamefont
  {M.}~\bibnamefont {Ramsey-Musolf}}, \emph {et~al.},\ }\bibfield  {title}
  {\enquote {\bibinfo {title} {Searches for new sources of cp violation using
  molecules as quantum sensors},}\ }\href@noop {} {\bibfield  {journal}
  {\bibinfo  {journal} {arXiv preprint arXiv:2010.08709}\ } (\bibinfo {year}
  {2020})}\BibitemShut {NoStop}%
\bibitem [{\citenamefont {Cairncross}\ \emph {et~al.}(2017)\citenamefont
  {Cairncross}, \citenamefont {Gresh}, \citenamefont {Grau}, \citenamefont
  {Cossel}, \citenamefont {Roussy}, \citenamefont {Ni}, \citenamefont {Zhou},
  \citenamefont {Ye},\ and\ \citenamefont {Cornell}}]{Cornell:2017}%
  \BibitemOpen
  \bibfield  {author} {\bibinfo {author} {\bibfnamefont {W.~B.}\ \bibnamefont
  {Cairncross}}, \bibinfo {author} {\bibfnamefont {D.~N.}\ \bibnamefont
  {Gresh}}, \bibinfo {author} {\bibfnamefont {M.}~\bibnamefont {Grau}},
  \bibinfo {author} {\bibfnamefont {K.~C.}\ \bibnamefont {Cossel}}, \bibinfo
  {author} {\bibfnamefont {T.~S.}\ \bibnamefont {Roussy}}, \bibinfo {author}
  {\bibfnamefont {Y.}~\bibnamefont {Ni}}, \bibinfo {author} {\bibfnamefont
  {Y.}~\bibnamefont {Zhou}}, \bibinfo {author} {\bibfnamefont {J.}~\bibnamefont
  {Ye}},\ and\ \bibinfo {author} {\bibfnamefont {E.~A.}\ \bibnamefont
  {Cornell}},\ }\bibfield  {title} {\enquote {\bibinfo {title} {Precision
  measurement of the electron's electric dipole moment using trapped molecular
  ions},}\ }\href@noop {} {\bibfield  {journal} {\bibinfo  {journal} {Phys.
  Rev. Lett.}\ }\textbf {\bibinfo {volume} {119}},\ \bibinfo {pages} {153001}
  (\bibinfo {year} {2017})}\BibitemShut {NoStop}%
\bibitem [{\citenamefont {Petrov}(2018)}]{Petrov:18}%
  \BibitemOpen
  \bibfield  {author} {\bibinfo {author} {\bibfnamefont {A.~N.}\ \bibnamefont
  {Petrov}},\ }\bibfield  {title} {\enquote {\bibinfo {title} {{Systematic
  effects in the ${\mathrm{HfF}}^{+}$-ion experiment to search for the electron
  electric dipole moment}},}\ }\href@noop {} {\bibfield  {journal} {\bibinfo
  {journal} {Phys. Rev. A}\ }\textbf {\bibinfo {volume} {97}},\ \bibinfo
  {pages} {052504} (\bibinfo {year} {2018})}\BibitemShut {NoStop}%
\bibitem [{\citenamefont {Alarcon}\ \emph {et~al.}(2022)\citenamefont
  {Alarcon}, \citenamefont {Alexander}, \citenamefont {Anastassopoulos},
  \citenamefont {Aoki}, \citenamefont {Baartman}, \citenamefont {Baeßler},
  \citenamefont {Bartoszek}, \citenamefont {Beck}, \citenamefont {Bedeschi},
  \citenamefont {Berger}, \citenamefont {Berz}, \citenamefont {Bethlem},
  \citenamefont {Bhattacharya}, \citenamefont {Blaskiewicz}, \citenamefont
  {Blum}, \citenamefont {Bowcock}, \citenamefont {Borschevsky}, \citenamefont
  {Brown}, \citenamefont {Budker}, \citenamefont {Burdin}, \citenamefont
  {Casey}, \citenamefont {Casse}, \citenamefont {Cantatore}, \citenamefont
  {Cheng}, \citenamefont {Chupp}, \citenamefont {Cianciolo}, \citenamefont
  {Cirigliano}, \citenamefont {Clayton}, \citenamefont {Crawford},
  \citenamefont {Das}, \citenamefont {Davoudiasl}, \citenamefont {de~Vries},
  \citenamefont {DeMille}, \citenamefont {Denisov}, \citenamefont {Diwan},
  \citenamefont {Doyle}, \citenamefont {Engel}, \citenamefont {Fanourakis},
  \citenamefont {Fatemi}, \citenamefont {Filippone}, \citenamefont {Flambaum},
  \citenamefont {Fleig}, \citenamefont {Fomin}, \citenamefont {Fischer},
  \citenamefont {Gabrielse}, \citenamefont {Ruiz}, \citenamefont {Gardikiotis},
  \citenamefont {Gatti}, \citenamefont {Geraci}, \citenamefont {Gooding},
  \citenamefont {Golub}, \citenamefont {Graham}, \citenamefont {Gray},
  \citenamefont {Griffith}, \citenamefont {Haciomeroglu}, \citenamefont
  {Gwinner}, \citenamefont {Hoekstra}, \citenamefont {Hoffstaetter},
  \citenamefont {Huang}, \citenamefont {Hutzler}, \citenamefont {Incagli},
  \citenamefont {Ito}, \citenamefont {Izubuchi}, \citenamefont {Jayich},
  \citenamefont {Jeong}, \citenamefont {Kaplan}, \citenamefont {Karuza},
  \citenamefont {Kawall}, \citenamefont {Kim}, \citenamefont {Koop},
  \citenamefont {Korsch}, \citenamefont {Korobkina}, \citenamefont {Lebedev},
  \citenamefont {Lee}, \citenamefont {Lee}, \citenamefont {Lehnert},
  \citenamefont {Leung}, \citenamefont {Liu}, \citenamefont {Long},
  \citenamefont {Lusiani}, \citenamefont {Marciano}, \citenamefont {Maroudas},
  \citenamefont {Matlashov}, \citenamefont {Matsumoto}, \citenamefont
  {Mawhorter}, \citenamefont {Meot}, \citenamefont {Mereghetti}, \citenamefont
  {Miller}, \citenamefont {Morse}, \citenamefont {Mott}, \citenamefont
  {Omarov}, \citenamefont {Orozco}, \citenamefont {O'Shaughnessy},
  \citenamefont {Ozben}, \citenamefont {Park}, \citenamefont {Pattie},
  \citenamefont {Petrov}, \citenamefont {Piacentino}, \citenamefont {Plaster},
  \citenamefont {Podobedov}, \citenamefont {Poelker}, \citenamefont {Pocanic},
  \citenamefont {Prasannaa}, \citenamefont {Price}, \citenamefont
  {Ramsey-Musolf}, \citenamefont {Raparia}, \citenamefont {Rajendran},
  \citenamefont {Reece}, \citenamefont {Reid}, \citenamefont {Rescia},
  \citenamefont {Ritz}, \citenamefont {Roberts}, \citenamefont {Safronova},
  \citenamefont {Sakemi}, \citenamefont {Schmidt-Wellenburg}, \citenamefont
  {Shindler}, \citenamefont {Semertzidis}, \citenamefont {Silenko},
  \citenamefont {Singh}, \citenamefont {Skripnikov}, \citenamefont {Soni},
  \citenamefont {Stephenson}, \citenamefont {Suleiman}, \citenamefont {Sunaga},
  \citenamefont {Syphers}, \citenamefont {Syritsyn}, \citenamefont {Tarbutt},
  \citenamefont {Thoerngren}, \citenamefont {Timmermans}, \citenamefont
  {Tishchenko}, \citenamefont {Titov}, \citenamefont {Tsoupas}, \citenamefont
  {Tzamarias}, \citenamefont {Variola}, \citenamefont {Venanzoni},
  \citenamefont {Vilella}, \citenamefont {Vossebeld}, \citenamefont {Winter},
  \citenamefont {Won}, \citenamefont {Zelenski}, \citenamefont {Zelevinsky},
  \citenamefont {Zhou},\ and\ \citenamefont {Zioutas}}]{whitepaper}%
  \BibitemOpen
  \bibfield  {author} {\bibinfo {author} {\bibfnamefont {R.}~\bibnamefont
  {Alarcon}}, \bibinfo {author} {\bibfnamefont {J.}~\bibnamefont {Alexander}},
  \bibinfo {author} {\bibfnamefont {V.}~\bibnamefont {Anastassopoulos}},
  \bibinfo {author} {\bibfnamefont {T.}~\bibnamefont {Aoki}}, \bibinfo {author}
  {\bibfnamefont {R.}~\bibnamefont {Baartman}}, \bibinfo {author}
  {\bibfnamefont {S.}~\bibnamefont {Baeßler}}, \bibinfo {author}
  {\bibfnamefont {L.}~\bibnamefont {Bartoszek}}, \bibinfo {author}
  {\bibfnamefont {D.~H.}\ \bibnamefont {Beck}}, \bibinfo {author}
  {\bibfnamefont {F.}~\bibnamefont {Bedeschi}}, \bibinfo {author}
  {\bibfnamefont {R.}~\bibnamefont {Berger}}, \bibinfo {author} {\bibfnamefont
  {M.}~\bibnamefont {Berz}}, \bibinfo {author} {\bibfnamefont {H.~L.}\
  \bibnamefont {Bethlem}}, \bibinfo {author} {\bibfnamefont {T.}~\bibnamefont
  {Bhattacharya}}, \bibinfo {author} {\bibfnamefont {M.}~\bibnamefont
  {Blaskiewicz}}, \bibinfo {author} {\bibfnamefont {T.}~\bibnamefont {Blum}},
  \bibinfo {author} {\bibfnamefont {T.}~\bibnamefont {Bowcock}}, \bibinfo
  {author} {\bibfnamefont {A.}~\bibnamefont {Borschevsky}}, \bibinfo {author}
  {\bibfnamefont {K.}~\bibnamefont {Brown}}, \bibinfo {author} {\bibfnamefont
  {D.}~\bibnamefont {Budker}}, \bibinfo {author} {\bibfnamefont
  {S.}~\bibnamefont {Burdin}}, \bibinfo {author} {\bibfnamefont {B.~C.}\
  \bibnamefont {Casey}}, \bibinfo {author} {\bibfnamefont {G.}~\bibnamefont
  {Casse}}, \bibinfo {author} {\bibfnamefont {G.}~\bibnamefont {Cantatore}},
  \bibinfo {author} {\bibfnamefont {L.}~\bibnamefont {Cheng}}, \bibinfo
  {author} {\bibfnamefont {T.}~\bibnamefont {Chupp}}, \bibinfo {author}
  {\bibfnamefont {V.}~\bibnamefont {Cianciolo}}, \bibinfo {author}
  {\bibfnamefont {V.}~\bibnamefont {Cirigliano}}, \bibinfo {author}
  {\bibfnamefont {S.~M.}\ \bibnamefont {Clayton}}, \bibinfo {author}
  {\bibfnamefont {C.}~\bibnamefont {Crawford}}, \bibinfo {author}
  {\bibfnamefont {B.~P.}\ \bibnamefont {Das}}, \bibinfo {author} {\bibfnamefont
  {H.}~\bibnamefont {Davoudiasl}}, \bibinfo {author} {\bibfnamefont
  {J.}~\bibnamefont {de~Vries}}, \bibinfo {author} {\bibfnamefont
  {D.}~\bibnamefont {DeMille}}, \bibinfo {author} {\bibfnamefont
  {D.}~\bibnamefont {Denisov}}, \bibinfo {author} {\bibfnamefont {M.~V.}\
  \bibnamefont {Diwan}}, \bibinfo {author} {\bibfnamefont {J.~M.}\ \bibnamefont
  {Doyle}}, \bibinfo {author} {\bibfnamefont {J.}~\bibnamefont {Engel}},
  \bibinfo {author} {\bibfnamefont {G.}~\bibnamefont {Fanourakis}}, \bibinfo
  {author} {\bibfnamefont {R.}~\bibnamefont {Fatemi}}, \bibinfo {author}
  {\bibfnamefont {B.~W.}\ \bibnamefont {Filippone}}, \bibinfo {author}
  {\bibfnamefont {V.~V.}\ \bibnamefont {Flambaum}}, \bibinfo {author}
  {\bibfnamefont {T.}~\bibnamefont {Fleig}}, \bibinfo {author} {\bibfnamefont
  {N.}~\bibnamefont {Fomin}}, \bibinfo {author} {\bibfnamefont
  {W.}~\bibnamefont {Fischer}}, \bibinfo {author} {\bibfnamefont
  {G.}~\bibnamefont {Gabrielse}}, \bibinfo {author} {\bibfnamefont {R.~F.~G.}\
  \bibnamefont {Ruiz}}, \bibinfo {author} {\bibfnamefont {A.}~\bibnamefont
  {Gardikiotis}}, \bibinfo {author} {\bibfnamefont {C.}~\bibnamefont {Gatti}},
  \bibinfo {author} {\bibfnamefont {A.}~\bibnamefont {Geraci}}, \bibinfo
  {author} {\bibfnamefont {J.}~\bibnamefont {Gooding}}, \bibinfo {author}
  {\bibfnamefont {B.}~\bibnamefont {Golub}}, \bibinfo {author} {\bibfnamefont
  {P.}~\bibnamefont {Graham}}, \bibinfo {author} {\bibfnamefont
  {F.}~\bibnamefont {Gray}}, \bibinfo {author} {\bibfnamefont {W.~C.}\
  \bibnamefont {Griffith}}, \bibinfo {author} {\bibfnamefont {S.}~\bibnamefont
  {Haciomeroglu}}, \bibinfo {author} {\bibfnamefont {G.}~\bibnamefont
  {Gwinner}}, \bibinfo {author} {\bibfnamefont {S.}~\bibnamefont {Hoekstra}},
  \bibinfo {author} {\bibfnamefont {G.~H.}\ \bibnamefont {Hoffstaetter}},
  \bibinfo {author} {\bibfnamefont {H.}~\bibnamefont {Huang}}, \bibinfo
  {author} {\bibfnamefont {N.~R.}\ \bibnamefont {Hutzler}}, \bibinfo {author}
  {\bibfnamefont {M.}~\bibnamefont {Incagli}}, \bibinfo {author} {\bibfnamefont
  {T.~M.}\ \bibnamefont {Ito}}, \bibinfo {author} {\bibfnamefont
  {T.}~\bibnamefont {Izubuchi}}, \bibinfo {author} {\bibfnamefont {A.~M.}\
  \bibnamefont {Jayich}}, \bibinfo {author} {\bibfnamefont {H.}~\bibnamefont
  {Jeong}}, \bibinfo {author} {\bibfnamefont {D.}~\bibnamefont {Kaplan}},
  \bibinfo {author} {\bibfnamefont {M.}~\bibnamefont {Karuza}}, \bibinfo
  {author} {\bibfnamefont {D.}~\bibnamefont {Kawall}}, \bibinfo {author}
  {\bibfnamefont {O.}~\bibnamefont {Kim}}, \bibinfo {author} {\bibfnamefont
  {I.}~\bibnamefont {Koop}}, \bibinfo {author} {\bibfnamefont {W.}~\bibnamefont
  {Korsch}}, \bibinfo {author} {\bibfnamefont {E.}~\bibnamefont {Korobkina}},
  \bibinfo {author} {\bibfnamefont {V.}~\bibnamefont {Lebedev}}, \bibinfo
  {author} {\bibfnamefont {J.}~\bibnamefont {Lee}}, \bibinfo {author}
  {\bibfnamefont {S.}~\bibnamefont {Lee}}, \bibinfo {author} {\bibfnamefont
  {R.}~\bibnamefont {Lehnert}}, \bibinfo {author} {\bibfnamefont {K.~K.~H.}\
  \bibnamefont {Leung}}, \bibinfo {author} {\bibfnamefont {C.-Y.}\ \bibnamefont
  {Liu}}, \bibinfo {author} {\bibfnamefont {J.}~\bibnamefont {Long}}, \bibinfo
  {author} {\bibfnamefont {A.}~\bibnamefont {Lusiani}}, \bibinfo {author}
  {\bibfnamefont {W.~J.}\ \bibnamefont {Marciano}}, \bibinfo {author}
  {\bibfnamefont {M.}~\bibnamefont {Maroudas}}, \bibinfo {author}
  {\bibfnamefont {A.}~\bibnamefont {Matlashov}}, \bibinfo {author}
  {\bibfnamefont {N.}~\bibnamefont {Matsumoto}}, \bibinfo {author}
  {\bibfnamefont {R.}~\bibnamefont {Mawhorter}}, \bibinfo {author}
  {\bibfnamefont {F.}~\bibnamefont {Meot}}, \bibinfo {author} {\bibfnamefont
  {E.}~\bibnamefont {Mereghetti}}, \bibinfo {author} {\bibfnamefont {J.~P.}\
  \bibnamefont {Miller}}, \bibinfo {author} {\bibfnamefont {W.~M.}\
  \bibnamefont {Morse}}, \bibinfo {author} {\bibfnamefont {J.}~\bibnamefont
  {Mott}}, \bibinfo {author} {\bibfnamefont {Z.}~\bibnamefont {Omarov}},
  \bibinfo {author} {\bibfnamefont {L.~A.}\ \bibnamefont {Orozco}}, \bibinfo
  {author} {\bibfnamefont {C.~M.}\ \bibnamefont {O'Shaughnessy}}, \bibinfo
  {author} {\bibfnamefont {C.}~\bibnamefont {Ozben}}, \bibinfo {author}
  {\bibfnamefont {S.}~\bibnamefont {Park}}, \bibinfo {author} {\bibfnamefont
  {R.~W.}\ \bibnamefont {Pattie}}, \bibinfo {author} {\bibfnamefont {A.~N.}\
  \bibnamefont {Petrov}}, \bibinfo {author} {\bibfnamefont {G.~M.}\
  \bibnamefont {Piacentino}}, \bibinfo {author} {\bibfnamefont {B.~R.}\
  \bibnamefont {Plaster}}, \bibinfo {author} {\bibfnamefont {B.}~\bibnamefont
  {Podobedov}}, \bibinfo {author} {\bibfnamefont {M.}~\bibnamefont {Poelker}},
  \bibinfo {author} {\bibfnamefont {D.}~\bibnamefont {Pocanic}}, \bibinfo
  {author} {\bibfnamefont {V.~S.}\ \bibnamefont {Prasannaa}}, \bibinfo {author}
  {\bibfnamefont {J.}~\bibnamefont {Price}}, \bibinfo {author} {\bibfnamefont
  {M.~J.}\ \bibnamefont {Ramsey-Musolf}}, \bibinfo {author} {\bibfnamefont
  {D.}~\bibnamefont {Raparia}}, \bibinfo {author} {\bibfnamefont
  {S.}~\bibnamefont {Rajendran}}, \bibinfo {author} {\bibfnamefont
  {M.}~\bibnamefont {Reece}}, \bibinfo {author} {\bibfnamefont
  {A.}~\bibnamefont {Reid}}, \bibinfo {author} {\bibfnamefont {S.}~\bibnamefont
  {Rescia}}, \bibinfo {author} {\bibfnamefont {A.}~\bibnamefont {Ritz}},
  \bibinfo {author} {\bibfnamefont {B.~L.}\ \bibnamefont {Roberts}}, \bibinfo
  {author} {\bibfnamefont {M.~S.}\ \bibnamefont {Safronova}}, \bibinfo {author}
  {\bibfnamefont {Y.}~\bibnamefont {Sakemi}}, \bibinfo {author} {\bibfnamefont
  {P.}~\bibnamefont {Schmidt-Wellenburg}}, \bibinfo {author} {\bibfnamefont
  {A.}~\bibnamefont {Shindler}}, \bibinfo {author} {\bibfnamefont {Y.~K.}\
  \bibnamefont {Semertzidis}}, \bibinfo {author} {\bibfnamefont
  {A.}~\bibnamefont {Silenko}}, \bibinfo {author} {\bibfnamefont {J.~T.}\
  \bibnamefont {Singh}}, \bibinfo {author} {\bibfnamefont {L.~V.}\ \bibnamefont
  {Skripnikov}}, \bibinfo {author} {\bibfnamefont {A.}~\bibnamefont {Soni}},
  \bibinfo {author} {\bibfnamefont {E.}~\bibnamefont {Stephenson}}, \bibinfo
  {author} {\bibfnamefont {R.}~\bibnamefont {Suleiman}}, \bibinfo {author}
  {\bibfnamefont {A.}~\bibnamefont {Sunaga}}, \bibinfo {author} {\bibfnamefont
  {M.}~\bibnamefont {Syphers}}, \bibinfo {author} {\bibfnamefont
  {S.}~\bibnamefont {Syritsyn}}, \bibinfo {author} {\bibfnamefont {M.~R.}\
  \bibnamefont {Tarbutt}}, \bibinfo {author} {\bibfnamefont {P.}~\bibnamefont
  {Thoerngren}}, \bibinfo {author} {\bibfnamefont {R.~G.~E.}\ \bibnamefont
  {Timmermans}}, \bibinfo {author} {\bibfnamefont {V.}~\bibnamefont
  {Tishchenko}}, \bibinfo {author} {\bibfnamefont {A.~V.}\ \bibnamefont
  {Titov}}, \bibinfo {author} {\bibfnamefont {N.}~\bibnamefont {Tsoupas}},
  \bibinfo {author} {\bibfnamefont {S.}~\bibnamefont {Tzamarias}}, \bibinfo
  {author} {\bibfnamefont {A.}~\bibnamefont {Variola}}, \bibinfo {author}
  {\bibfnamefont {G.}~\bibnamefont {Venanzoni}}, \bibinfo {author}
  {\bibfnamefont {E.}~\bibnamefont {Vilella}}, \bibinfo {author} {\bibfnamefont
  {J.}~\bibnamefont {Vossebeld}}, \bibinfo {author} {\bibfnamefont
  {P.}~\bibnamefont {Winter}}, \bibinfo {author} {\bibfnamefont
  {E.}~\bibnamefont {Won}}, \bibinfo {author} {\bibfnamefont {A.}~\bibnamefont
  {Zelenski}}, \bibinfo {author} {\bibfnamefont {T.}~\bibnamefont
  {Zelevinsky}}, \bibinfo {author} {\bibfnamefont {Y.}~\bibnamefont {Zhou}},\
  and\ \bibinfo {author} {\bibfnamefont {K.}~\bibnamefont {Zioutas}},\ }\href
  {https://doi.org/10.48550/ARXIV.2203.08103} {\enquote {\bibinfo {title}
  {Electric dipole moments and the search for new physics},}\ } (\bibinfo
  {year} {2022})\BibitemShut {NoStop}%
\bibitem [{\citenamefont {Isaev}\ and\ \citenamefont
  {Berger}(2016)}]{Isaev:16}%
  \BibitemOpen
  \bibfield  {author} {\bibinfo {author} {\bibfnamefont {T.~A.}\ \bibnamefont
  {Isaev}}\ and\ \bibinfo {author} {\bibfnamefont {R.}~\bibnamefont {Berger}},\
  }\bibfield  {title} {\enquote {\bibinfo {title} {Polyatomic candidates for
  cooling of molecules with lasers from simple theoretical concepts},}\ }\href
  {https://doi.org/10.1103/PhysRevLett.116.063006} {\bibfield  {journal}
  {\bibinfo  {journal} {Phys. Rev. Lett.}\ }\textbf {\bibinfo {volume} {116}},\
  \bibinfo {pages} {063006} (\bibinfo {year} {2016})}\BibitemShut {NoStop}%
\bibitem [{\citenamefont {Isaev}, \citenamefont {Zaitsevskii},\ and\
  \citenamefont {Eliav}(2017)}]{Isaev_2017}%
  \BibitemOpen
  \bibfield  {author} {\bibinfo {author} {\bibfnamefont {T.~A.}\ \bibnamefont
  {Isaev}}, \bibinfo {author} {\bibfnamefont {A.~V.}\ \bibnamefont
  {Zaitsevskii}},\ and\ \bibinfo {author} {\bibfnamefont {E.}~\bibnamefont
  {Eliav}},\ }\bibfield  {title} {\enquote {\bibinfo {title} {Laser-coolable
  polyatomic molecules with heavy nuclei},}\ }\href
  {https://doi.org/10.1088/1361-6455/aa8f34} {\bibfield  {journal} {\bibinfo
  {journal} {Journal of Physics B: Atomic, Molecular and Optical Physics}\
  }\textbf {\bibinfo {volume} {50}},\ \bibinfo {pages} {225101} (\bibinfo
  {year} {2017})}\BibitemShut {NoStop}%
\bibitem [{\citenamefont {Kozyryev}\ and\ \citenamefont
  {Hutzler}(2017)}]{Kozyryev:17}%
  \BibitemOpen
  \bibfield  {author} {\bibinfo {author} {\bibfnamefont {I.}~\bibnamefont
  {Kozyryev}}\ and\ \bibinfo {author} {\bibfnamefont {N.~R.}\ \bibnamefont
  {Hutzler}},\ }\bibfield  {title} {\enquote {\bibinfo {title} {Precision
  measurement of time-reversal symmetry violation with laser-cooled polyatomic
  molecules},}\ }\href {https://doi.org/10.1103/PhysRevLett.119.133002}
  {\bibfield  {journal} {\bibinfo  {journal} {Phys. Rev. Lett.}\ }\textbf
  {\bibinfo {volume} {119}},\ \bibinfo {pages} {133002} (\bibinfo {year}
  {2017})}\BibitemShut {NoStop}%
\bibitem [{\citenamefont {Hutzler}(2020)}]{hutzler2020polyatomic}%
  \BibitemOpen
  \bibfield  {author} {\bibinfo {author} {\bibfnamefont {N.~R.}\ \bibnamefont
  {Hutzler}},\ }\bibfield  {title} {\enquote {\bibinfo {title} {Polyatomic
  molecules as quantum sensors for fundamental physics},}\ }\href@noop {}
  {\bibfield  {journal} {\bibinfo  {journal} {Quantum Science and Technology}\
  }\textbf {\bibinfo {volume} {5}},\ \bibinfo {pages} {044011} (\bibinfo {year}
  {2020})}\BibitemShut {NoStop}%
\bibitem [{\citenamefont {Zakharova}(2022)}]{zakharova2022rotating}%
  \BibitemOpen
  \bibfield  {author} {\bibinfo {author} {\bibfnamefont {A.}~\bibnamefont
  {Zakharova}},\ }\bibfield  {title} {\enquote {\bibinfo {title} {Rotating and
  vibrating symmetric-top molecule raoch 3 in fundamental p, t-violation
  searches},}\ }\href@noop {} {\bibfield  {journal} {\bibinfo  {journal}
  {Physical Review A}\ }\textbf {\bibinfo {volume} {105}},\ \bibinfo {pages}
  {032811} (\bibinfo {year} {2022})}\BibitemShut {NoStop}%
\bibitem [{\citenamefont {Kozyryev}\ \emph {et~al.}(2017)\citenamefont
  {Kozyryev}, \citenamefont {Baum}, \citenamefont {Matsuda}, \citenamefont
  {Augenbraun}, \citenamefont {Anderegg}, \citenamefont {Sedlack},\ and\
  \citenamefont {Doyle}}]{kozyryev2017sisyphus}%
  \BibitemOpen
  \bibfield  {author} {\bibinfo {author} {\bibfnamefont {I.}~\bibnamefont
  {Kozyryev}}, \bibinfo {author} {\bibfnamefont {L.}~\bibnamefont {Baum}},
  \bibinfo {author} {\bibfnamefont {K.}~\bibnamefont {Matsuda}}, \bibinfo
  {author} {\bibfnamefont {B.~L.}\ \bibnamefont {Augenbraun}}, \bibinfo
  {author} {\bibfnamefont {L.}~\bibnamefont {Anderegg}}, \bibinfo {author}
  {\bibfnamefont {A.~P.}\ \bibnamefont {Sedlack}},\ and\ \bibinfo {author}
  {\bibfnamefont {J.~M.}\ \bibnamefont {Doyle}},\ }\bibfield  {title} {\enquote
  {\bibinfo {title} {Sisyphus laser cooling of a polyatomic molecule},}\
  }\href@noop {} {\bibfield  {journal} {\bibinfo  {journal} {Physical review
  letters}\ }\textbf {\bibinfo {volume} {118}},\ \bibinfo {pages} {173201}
  (\bibinfo {year} {2017})}\BibitemShut {NoStop}%
\bibitem [{\citenamefont {Baum}\ \emph {et~al.}(2020)\citenamefont {Baum},
  \citenamefont {Vilas}, \citenamefont {Hallas}, \citenamefont {Augenbraun},
  \citenamefont {Raval}, \citenamefont {Mitra},\ and\ \citenamefont
  {Doyle}}]{baum20201d}%
  \BibitemOpen
  \bibfield  {author} {\bibinfo {author} {\bibfnamefont {L.}~\bibnamefont
  {Baum}}, \bibinfo {author} {\bibfnamefont {N.~B.}\ \bibnamefont {Vilas}},
  \bibinfo {author} {\bibfnamefont {C.}~\bibnamefont {Hallas}}, \bibinfo
  {author} {\bibfnamefont {B.~L.}\ \bibnamefont {Augenbraun}}, \bibinfo
  {author} {\bibfnamefont {S.}~\bibnamefont {Raval}}, \bibinfo {author}
  {\bibfnamefont {D.}~\bibnamefont {Mitra}},\ and\ \bibinfo {author}
  {\bibfnamefont {J.~M.}\ \bibnamefont {Doyle}},\ }\bibfield  {title} {\enquote
  {\bibinfo {title} {1d magneto-optical trap of polyatomic molecules},}\
  }\href@noop {} {\bibfield  {journal} {\bibinfo  {journal} {Physical review
  letters}\ }\textbf {\bibinfo {volume} {124}},\ \bibinfo {pages} {133201}
  (\bibinfo {year} {2020})}\BibitemShut {NoStop}%
\bibitem [{\citenamefont {Mitra}\ \emph {et~al.}(2020)\citenamefont {Mitra},
  \citenamefont {Vilas}, \citenamefont {Hallas}, \citenamefont {Anderegg},
  \citenamefont {Augenbraun}, \citenamefont {Baum}, \citenamefont {Miller},
  \citenamefont {Raval},\ and\ \citenamefont {Doyle}}]{mitra2020direct}%
  \BibitemOpen
  \bibfield  {author} {\bibinfo {author} {\bibfnamefont {D.}~\bibnamefont
  {Mitra}}, \bibinfo {author} {\bibfnamefont {N.~B.}\ \bibnamefont {Vilas}},
  \bibinfo {author} {\bibfnamefont {C.}~\bibnamefont {Hallas}}, \bibinfo
  {author} {\bibfnamefont {L.}~\bibnamefont {Anderegg}}, \bibinfo {author}
  {\bibfnamefont {B.~L.}\ \bibnamefont {Augenbraun}}, \bibinfo {author}
  {\bibfnamefont {L.}~\bibnamefont {Baum}}, \bibinfo {author} {\bibfnamefont
  {C.}~\bibnamefont {Miller}}, \bibinfo {author} {\bibfnamefont
  {S.}~\bibnamefont {Raval}},\ and\ \bibinfo {author} {\bibfnamefont {J.~M.}\
  \bibnamefont {Doyle}},\ }\bibfield  {title} {\enquote {\bibinfo {title}
  {Direct laser cooling of a symmetric top molecule},}\ }\href@noop {}
  {\bibfield  {journal} {\bibinfo  {journal} {Science}\ }\textbf {\bibinfo
  {volume} {369}},\ \bibinfo {pages} {1366--1369} (\bibinfo {year}
  {2020})}\BibitemShut {NoStop}%
\bibitem [{\citenamefont {Steimle}\ \emph {et~al.}(2019)\citenamefont
  {Steimle}, \citenamefont {Linton}, \citenamefont {Mengesha}, \citenamefont
  {Bai},\ and\ \citenamefont {Le}}]{steimle2019field}%
  \BibitemOpen
  \bibfield  {author} {\bibinfo {author} {\bibfnamefont {T.~C.}\ \bibnamefont
  {Steimle}}, \bibinfo {author} {\bibfnamefont {C.}~\bibnamefont {Linton}},
  \bibinfo {author} {\bibfnamefont {E.~T.}\ \bibnamefont {Mengesha}}, \bibinfo
  {author} {\bibfnamefont {X.}~\bibnamefont {Bai}},\ and\ \bibinfo {author}
  {\bibfnamefont {A.~T.}\ \bibnamefont {Le}},\ }\bibfield  {title} {\enquote
  {\bibinfo {title} {{Field-free, Stark, and Zeeman spectroscopy of the
  $\tilde{A}^2 \Pi_{1/2} - \tilde{X}^2\Sigma^+$ transition of ytterbium
  monohydroxide}},}\ }\href@noop {} {\bibfield  {journal} {\bibinfo  {journal}
  {Physical Review A}\ }\textbf {\bibinfo {volume} {100}},\ \bibinfo {pages}
  {052509} (\bibinfo {year} {2019})}\BibitemShut {NoStop}%
\bibitem [{\citenamefont {Augenbraun}\ \emph {et~al.}(2020)\citenamefont
  {Augenbraun}, \citenamefont {Lasner}, \citenamefont {Frenett}, \citenamefont
  {Sawaoka}, \citenamefont {Miller}, \citenamefont {Steimle},\ and\
  \citenamefont {Doyle}}]{augenbraun2020laser}%
  \BibitemOpen
  \bibfield  {author} {\bibinfo {author} {\bibfnamefont {B.~L.}\ \bibnamefont
  {Augenbraun}}, \bibinfo {author} {\bibfnamefont {Z.~D.}\ \bibnamefont
  {Lasner}}, \bibinfo {author} {\bibfnamefont {A.}~\bibnamefont {Frenett}},
  \bibinfo {author} {\bibfnamefont {H.}~\bibnamefont {Sawaoka}}, \bibinfo
  {author} {\bibfnamefont {C.}~\bibnamefont {Miller}}, \bibinfo {author}
  {\bibfnamefont {T.~C.}\ \bibnamefont {Steimle}},\ and\ \bibinfo {author}
  {\bibfnamefont {J.~M.}\ \bibnamefont {Doyle}},\ }\bibfield  {title} {\enquote
  {\bibinfo {title} {Laser-cooled polyatomic molecules for improved electron
  electric dipole moment searches},}\ }\href@noop {} {\bibfield  {journal}
  {\bibinfo  {journal} {New Journal of Physics}\ }\textbf {\bibinfo {volume}
  {22}},\ \bibinfo {pages} {022003} (\bibinfo {year} {2020})}\BibitemShut
  {NoStop}%
\bibitem [{\citenamefont {Vilas}\ \emph {et~al.}(2022)\citenamefont {Vilas},
  \citenamefont {Hallas}, \citenamefont {Anderegg}, \citenamefont {Robichaud},
  \citenamefont {Winnicki}, \citenamefont {Mitra},\ and\ \citenamefont
  {Doyle}}]{vilas2022magneto}%
  \BibitemOpen
  \bibfield  {author} {\bibinfo {author} {\bibfnamefont {N.~B.}\ \bibnamefont
  {Vilas}}, \bibinfo {author} {\bibfnamefont {C.}~\bibnamefont {Hallas}},
  \bibinfo {author} {\bibfnamefont {L.}~\bibnamefont {Anderegg}}, \bibinfo
  {author} {\bibfnamefont {P.}~\bibnamefont {Robichaud}}, \bibinfo {author}
  {\bibfnamefont {A.}~\bibnamefont {Winnicki}}, \bibinfo {author}
  {\bibfnamefont {D.}~\bibnamefont {Mitra}},\ and\ \bibinfo {author}
  {\bibfnamefont {J.~M.}\ \bibnamefont {Doyle}},\ }\bibfield  {title} {\enquote
  {\bibinfo {title} {Magneto-optical trapping and sub-doppler cooling of a
  polyatomic molecule},}\ }\href@noop {} {\bibfield  {journal} {\bibinfo
  {journal} {Nature}\ }\textbf {\bibinfo {volume} {606}},\ \bibinfo {pages}
  {70--74} (\bibinfo {year} {2022})}\BibitemShut {NoStop}%
\bibitem [{\citenamefont {Kozlov}\ and\ \citenamefont
  {Labzowsky}(1995)}]{KozlovLabzowsky1995}%
  \BibitemOpen
  \bibfield  {author} {\bibinfo {author} {\bibfnamefont {M.~G.}\ \bibnamefont
  {Kozlov}}\ and\ \bibinfo {author} {\bibfnamefont {L.~N.}\ \bibnamefont
  {Labzowsky}},\ }\bibfield  {title} {\enquote {\bibinfo {title} {Parity
  violation effects in diatomics},}\ }\href
  {https://doi.org/10.1088/0953-4075/28/10/008} {\bibfield  {journal} {\bibinfo
   {journal} {Journal of Physics B: Atomic, Molecular and Optical Physics}\
  }\textbf {\bibinfo {volume} {28}},\ \bibinfo {pages} {1933} (\bibinfo {year}
  {1995})}\BibitemShut {NoStop}%
\bibitem [{\citenamefont {Titov}\ \emph {et~al.}(2006)\citenamefont {Titov},
  \citenamefont {Mosyagin}, \citenamefont {Petrov}, \citenamefont {Isaev},\
  and\ \citenamefont {DeMille}}]{titov2006d}%
  \BibitemOpen
  \bibfield  {author} {\bibinfo {author} {\bibfnamefont {A.}~\bibnamefont
  {Titov}}, \bibinfo {author} {\bibfnamefont {N.}~\bibnamefont {Mosyagin}},
  \bibinfo {author} {\bibfnamefont {A.}~\bibnamefont {Petrov}}, \bibinfo
  {author} {\bibfnamefont {T.}~\bibnamefont {Isaev}},\ and\ \bibinfo {author}
  {\bibfnamefont {D.}~\bibnamefont {DeMille}},\ }\bibfield  {title} {\enquote
  {\bibinfo {title} {P, t-parity violation effects in polar heavy-atom
  molecules},}\ }in\ \href@noop {} {\emph {\bibinfo {booktitle} {Recent
  Advances in the Theory of Chemical and Physical Systems}}}\ (\bibinfo
  {publisher} {Springer},\ \bibinfo {year} {2006})\ pp.\ \bibinfo {pages}
  {253--283}\BibitemShut {NoStop}%
\bibitem [{\citenamefont {Safronova}\ \emph {et~al.}(2018)\citenamefont
  {Safronova}, \citenamefont {Budker}, \citenamefont {DeMille}, \citenamefont
  {Kimball}, \citenamefont {Derevianko},\ and\ \citenamefont
  {Clark}}]{Safronova2017}%
  \BibitemOpen
  \bibfield  {author} {\bibinfo {author} {\bibfnamefont {M.~S.}\ \bibnamefont
  {Safronova}}, \bibinfo {author} {\bibfnamefont {D.}~\bibnamefont {Budker}},
  \bibinfo {author} {\bibfnamefont {D.}~\bibnamefont {DeMille}}, \bibinfo
  {author} {\bibfnamefont {D.~F.~J.}\ \bibnamefont {Kimball}}, \bibinfo
  {author} {\bibfnamefont {A.}~\bibnamefont {Derevianko}},\ and\ \bibinfo
  {author} {\bibfnamefont {C.~W.}\ \bibnamefont {Clark}},\ }\bibfield  {title}
  {\enquote {\bibinfo {title} {{Search for New Physics with Atoms and
  Molecules}},}\ }\href {https://doi.org/10.1103/RevModPhys.90.025008}
  {\bibfield  {journal} {\bibinfo  {journal} {Rev. Mod. Phys.}\ }\textbf
  {\bibinfo {volume} {90}},\ \bibinfo {pages} {025008} (\bibinfo {year}
  {2018})},\ \Eprint {https://arxiv.org/abs/1710.01833} {arXiv:1710.01833
  [physics.atom-ph]} \BibitemShut {NoStop}%
\bibitem [{\citenamefont {Zakharova}\ and\ \citenamefont
  {Petrov}(2021)}]{ourRaOH}%
  \BibitemOpen
  \bibfield  {author} {\bibinfo {author} {\bibfnamefont {A.}~\bibnamefont
  {Zakharova}}\ and\ \bibinfo {author} {\bibfnamefont {A.}~\bibnamefont
  {Petrov}},\ }\bibfield  {title} {\enquote {\bibinfo {title}
  {{$\mathcal{P}$,$\mathcal{T}$-odd effects for the RaOH molecule in the
  excited vibrational state}},}\ }\href
  {https://doi.org/10.1103/PhysRevA.103.032819} {\bibfield  {journal} {\bibinfo
   {journal} {Phys. Rev. A}\ }\textbf {\bibinfo {volume} {103}},\ \bibinfo
  {pages} {032819} (\bibinfo {year} {2021})},\ \Eprint
  {https://arxiv.org/abs/2012.08427} {arXiv:2012.08427 [physics.atom-ph]}
  \BibitemShut {NoStop}%
\bibitem [{DIR()}]{DIRAC19}%
  \BibitemOpen
  \href@noop {} {}\bibinfo {note} {{DIRAC}, a relativistic ab initio electronic
  structure program, Release {DIRAC19} (2019), written by A.~S.~P.~Gomes,
  T.~Saue, L.~Visscher, H.~J.~{\relax Aa}.~Jensen, and R.~Bast, with
  contributions from I.~A.~Aucar, V.~Bakken, K.~G.~Dyall, S.~Dubillard,
  U.~Ekstr{\"o}m, E.~Eliav, T.~Enevoldsen, E.~Fa{\ss}hauer, T.~Fleig,
  O.~Fossgaard, L.~Halbert, E.~D.~Hedeg{\aa}rd, B.~Heimlich--Paris,
  T.~Helgaker, J.~Henriksson, M.~Ilia{\v{s}}, Ch.~R.~Jacob, S.~Knecht,
  S.~Komorovsk{\'y}, O.~Kullie, J.~K.~L{\ae}rdahl, C.~V.~Larsen, Y.~S.~Lee,
  H.~S.~Nataraj, M.~K.~Nayak, P.~Norman, G.~Olejniczak, J.~Olsen,
  J.~M.~H.~Olsen, Y.~C.~Park, J.~K.~Pedersen, M.~Pernpointner, R.~di~Remigio,
  K.~Ruud, P.~Sa{\l}ek, B.~Schimmelpfennig, B.~Senjean, A.~Shee, J.~Sikkema,
  A.~J.~Thorvaldsen, J.~Thyssen, J.~van~Stralen, M.~L.~Vidal, S.~Villaume,
  O.~Visser, T.~Winther, and S.~Yamamoto (available at
  http://dx.doi.org/10.5281/zenodo.3572669, see also
  http://www.diracprogram.org)}\BibitemShut {NoStop}%
\bibitem [{\citenamefont {{URL:
  http://www.qchem.pnpi.spb.ru/Basis/~}}()}]{QCPNPI:Basis}%
  \BibitemOpen
  \bibfield  {author} {\bibinfo {author} {\bibnamefont {{URL:
  http://www.qchem.pnpi.spb.ru/Basis/~}}},\ }\href@noop {} {}\bibinfo {note}
  {~{GRECPs} and basis sets}\BibitemShut {NoStop}%
\bibitem [{\citenamefont {Titov}\ and\ \citenamefont
  {Mosyagin}(1999)}]{titov1999generalized}%
  \BibitemOpen
  \bibfield  {author} {\bibinfo {author} {\bibfnamefont {A.}~\bibnamefont
  {Titov}}\ and\ \bibinfo {author} {\bibfnamefont {N.}~\bibnamefont
  {Mosyagin}},\ }\bibfield  {title} {\enquote {\bibinfo {title} {Generalized
  relativistic effective core potential: Theoretical grounds},}\ }\href@noop {}
  {\bibfield  {journal} {\bibinfo  {journal} {International journal of quantum
  chemistry}\ }\textbf {\bibinfo {volume} {71}},\ \bibinfo {pages} {359--401}
  (\bibinfo {year} {1999})}\BibitemShut {NoStop}%
\bibitem [{\citenamefont {Mosyagin}, \citenamefont {Zaitsevskii},\ and\
  \citenamefont {Titov}(2010)}]{mosyagin2010shape}%
  \BibitemOpen
  \bibfield  {author} {\bibinfo {author} {\bibfnamefont {N.~S.}\ \bibnamefont
  {Mosyagin}}, \bibinfo {author} {\bibfnamefont {A.}~\bibnamefont
  {Zaitsevskii}},\ and\ \bibinfo {author} {\bibfnamefont {A.~V.}\ \bibnamefont
  {Titov}},\ }\bibfield  {title} {\enquote {\bibinfo {title} {Shape-consistent
  relativistic effective potentials of small atomic cores},}\ }\href@noop {}
  {\bibfield  {journal} {\bibinfo  {journal} {International Review of Atomic
  and Molecular Physics}\ }\textbf {\bibinfo {volume} {1}},\ \bibinfo {pages}
  {63--72} (\bibinfo {year} {2010})}\BibitemShut {NoStop}%
\bibitem [{\citenamefont {Mosyagin}\ \emph {et~al.}(2016)\citenamefont
  {Mosyagin}, \citenamefont {Zaitsevskii}, \citenamefont {Skripnikov},\ and\
  \citenamefont {Titov}}]{mosyagin2016generalized}%
  \BibitemOpen
  \bibfield  {author} {\bibinfo {author} {\bibfnamefont {N.~S.}\ \bibnamefont
  {Mosyagin}}, \bibinfo {author} {\bibfnamefont {A.~V.}\ \bibnamefont
  {Zaitsevskii}}, \bibinfo {author} {\bibfnamefont {L.~V.}\ \bibnamefont
  {Skripnikov}},\ and\ \bibinfo {author} {\bibfnamefont {A.~V.}\ \bibnamefont
  {Titov}},\ }\bibfield  {title} {\enquote {\bibinfo {title} {Generalized
  relativistic effective core potentials for actinides},}\ }\href@noop {}
  {\bibfield  {journal} {\bibinfo  {journal} {International Journal of Quantum
  Chemistry}\ }\textbf {\bibinfo {volume} {116}},\ \bibinfo {pages} {301--315}
  (\bibinfo {year} {2016})}\BibitemShut {NoStop}%
\bibitem [{\citenamefont {McGuire}\ and\ \citenamefont
  {Kouri}(1974)}]{mcguire1974quantum}%
  \BibitemOpen
  \bibfield  {author} {\bibinfo {author} {\bibfnamefont {P.}~\bibnamefont
  {McGuire}}\ and\ \bibinfo {author} {\bibfnamefont {D.~J.}\ \bibnamefont
  {Kouri}},\ }\bibfield  {title} {\enquote {\bibinfo {title} {{Quantum
  mechanical close coupling approach to molecular collisions. $j_z$-conserving
  coupled states approximation}},}\ }\href@noop {} {\bibfield  {journal}
  {\bibinfo  {journal} {The Journal of Chemical Physics}\ }\textbf {\bibinfo
  {volume} {60}},\ \bibinfo {pages} {2488--2499} (\bibinfo {year}
  {1974})}\BibitemShut {NoStop}%
\bibitem [{\citenamefont {Herzberg}(1966)}]{HerzbergBook}%
  \BibitemOpen
  \bibfield  {author} {\bibinfo {author} {\bibfnamefont {G.}~\bibnamefont
  {Herzberg}},\ }\href@noop {} {\emph {\bibinfo {title} {{Molecular spectra and
  molecular structure. Vol. 3: Electronic spectra and electronic structure of
  polyatomic molecules}}}}\ (\bibinfo  {publisher} {New York: Van Nostrand},\
  \bibinfo {year} {1966})\BibitemShut {NoStop}%
\bibitem [{\citenamefont {Melville}\ and\ \citenamefont
  {Coxon}(2001)}]{melville2001visible}%
  \BibitemOpen
  \bibfield  {author} {\bibinfo {author} {\bibfnamefont {T.~C.}\ \bibnamefont
  {Melville}}\ and\ \bibinfo {author} {\bibfnamefont {J.~A.}\ \bibnamefont
  {Coxon}},\ }\bibfield  {title} {\enquote {\bibinfo {title} {{The visible
  laser excitation spectrum of YbOH: The $\tilde{A}^2 \Pi-\tilde{X}^2 \Sigma^+$
  transition}},}\ }\href {https://doi.org/10.1063/1.1404145} {\bibfield
  {journal} {\bibinfo  {journal} {The Journal of Chemical Physics}\ }\textbf
  {\bibinfo {volume} {115}},\ \bibinfo {pages} {6974--6978} (\bibinfo {year}
  {2001})}\BibitemShut {NoStop}%
\bibitem [{\citenamefont {Zhang}\ \emph {et~al.}(2021)\citenamefont {Zhang},
  \citenamefont {Augenbraun}, \citenamefont {Lasner}, \citenamefont {Vilas},
  \citenamefont {Doyle},\ and\ \citenamefont {Cheng}}]{zhang2021accurate}%
  \BibitemOpen
  \bibfield  {author} {\bibinfo {author} {\bibfnamefont {C.}~\bibnamefont
  {Zhang}}, \bibinfo {author} {\bibfnamefont {B.~L.}\ \bibnamefont
  {Augenbraun}}, \bibinfo {author} {\bibfnamefont {Z.~D.}\ \bibnamefont
  {Lasner}}, \bibinfo {author} {\bibfnamefont {N.~B.}\ \bibnamefont {Vilas}},
  \bibinfo {author} {\bibfnamefont {J.~M.}\ \bibnamefont {Doyle}},\ and\
  \bibinfo {author} {\bibfnamefont {L.}~\bibnamefont {Cheng}},\ }\bibfield
  {title} {\enquote {\bibinfo {title} {Accurate prediction and measurement of
  vibronic branching ratios for laser cooling linear polyatomic molecules},}\
  }\href@noop {} {\bibfield  {journal} {\bibinfo  {journal} {The Journal of
  Chemical Physics}\ }\textbf {\bibinfo {volume} {155}},\ \bibinfo {pages}
  {091101} (\bibinfo {year} {2021})}\BibitemShut {NoStop}%
\bibitem [{\citenamefont {Mengesha}\ \emph {et~al.}(2020)\citenamefont
  {Mengesha}, \citenamefont {Le}, \citenamefont {Steimle}, \citenamefont
  {Cheng}, \citenamefont {Zhang}, \citenamefont {Augenbraun}, \citenamefont
  {Lasner},\ and\ \citenamefont {Doyle}}]{mengesha2020branching}%
  \BibitemOpen
  \bibfield  {author} {\bibinfo {author} {\bibfnamefont {E.~T.}\ \bibnamefont
  {Mengesha}}, \bibinfo {author} {\bibfnamefont {A.~T.}\ \bibnamefont {Le}},
  \bibinfo {author} {\bibfnamefont {T.~C.}\ \bibnamefont {Steimle}}, \bibinfo
  {author} {\bibfnamefont {L.}~\bibnamefont {Cheng}}, \bibinfo {author}
  {\bibfnamefont {C.}~\bibnamefont {Zhang}}, \bibinfo {author} {\bibfnamefont
  {B.~L.}\ \bibnamefont {Augenbraun}}, \bibinfo {author} {\bibfnamefont
  {Z.}~\bibnamefont {Lasner}},\ and\ \bibinfo {author} {\bibfnamefont
  {J.}~\bibnamefont {Doyle}},\ }\bibfield  {title} {\enquote {\bibinfo {title}
  {Branching ratios, radiative lifetimes, and transition dipole moments for
  yboh},}\ }\href@noop {} {\bibfield  {journal} {\bibinfo  {journal} {The
  Journal of Physical Chemistry A}\ }\textbf {\bibinfo {volume} {124}},\
  \bibinfo {pages} {3135--3148} (\bibinfo {year} {2020})}\BibitemShut {NoStop}%
\end{thebibliography}
%

\end{document}